\documentclass[a4paper,preprintnumbers,floatfix,twocolumn,aps,prb,unsortedaddress,superscriptaddress]{revtex4-2}

\usepackage[para]{threeparttable} 
\usepackage{amsmath}
\usepackage{graphicx} 
\usepackage{booktabs}
\usepackage{caption}
\usepackage{subcaption}

\usepackage{url}
\usepackage{hyperref}
\usepackage{miller}
\usepackage{bm}
\usepackage{enumitem}
\usepackage{gensymb}
\usepackage{color}
\usepackage{arydshln}
\usepackage{tabu}
\usepackage{xcolor}
\usepackage{tabularx}

\newcommand{\etal}{{\it et al. }}
\usepackage[paperwidth=210mm,paperheight=297mm,centering,hmargin=1.7cm,vmargin=2.5cm]{geometry}

\newcommand{\dashrule}[1][black]{%
  \color{#1}\rule[\dimexpr.5ex-.2pt]{4pt}{.4pt}\xleaders\hbox{\rule{4pt}{0pt}\rule[\dimexpr.5ex-.2pt]{4pt}{.4pt}}\hfill\kern0pt%
}
\newcommand{\rulecolor}[1]{%
  \def\CT@arc@{\color{#1}}%
}

\usepackage[utf8]{inputenc}

\renewcommand{\selectlanguage}[1]{} 

\begin{document}

\title{Effects of training machine-learning potentials for radiation damage simulations using different pseudopotentials}

\author{A. Fellman}
\thanks{Corresponding author}
\email{aslak.fellman@helsinki.fi}
\affiliation{Department of Physics, P.O. Box 43, FI-00014 University of Helsinki, Finland}
\author{J. Byggmästar}
\affiliation{Department of Physics, P.O. Box 43, FI-00014 University of Helsinki, Finland}
\author{F. Granberg}
\affiliation{Department of Physics, P.O. Box 43, FI-00014 University of Helsinki, Finland}
\author{F. Djurabekova}
\affiliation{Department of Physics, P.O. Box 43, FI-00014 University of Helsinki, Finland}
\author{K. Nordlund}
\affiliation{Department of Physics, P.O. Box 43, FI-00014 University of Helsinki, Finland}

\date{\today}

\begin{abstract}
Machine learning (ML) has become a commonplace approach in the development of interatomic potentials for molecular dynamics simulations, and its use also for radiation effect modelling is increasing. In this work, we investigate the effects of training ML potentials to density functional theory data calculated with different pseudopotentials in nickel. We look in detail at the differences that appear in radiation damage simulations. The use of a ''harder'' pseudopotential with semicore electrons has a direct impact on the short-range interactions, which in turn has implications on the radiation damage simulations. We find that despite these differences, the average threshold displacement energy is quite similar (40--50 eV for Ni). However, we find significant differences in the cumulative damage predicted by massively overlapping cascade simulations and compare them with  Rutherford Backscattering 
Spectroscopy/channeling experiments. We also investigate approaches to modify the repulsive pair interactions after training the potentials and discuss the feasibility of such approaches. 
\end{abstract}

\maketitle

\section{Introduction}
\label{sec:intro}

Molecular dynamics (MD) has been used to study radiation damage since the 1960s~\cite{nordlund_historical_2019}. During the past decades mostly classical analytical potentials have been used in MD simulations of radiation damage~\cite{nordlund_primary_2018}. In this period, it has been widely demonstrated that the accuracy of the interatomic potentials is key for reliable simulations of radiation damage. Extensive comparative studies have been performed using a wide range of classical analytical potentials, comparing the predictions from them~\cite{becquart_modelling_2021,de2021modelling}. It has been shown that the results can vary significantly based on the choice of interatomic potential. In the past, the repulsive interaction of the potentials were usually implemented via ''hardening'' of the potential, where the repulsive short-range pair interactions are joined to an existing potential fitted to near-equilibrium material properties. Later, more motivated approaches of joining the potentials have been suggested and applied, which for the most part rely on requiring a good transition from density functional theory (DFT) calculations towards the repulsive potentials \cite{nordlund2025repulsive}. It is also well known that the choice of pseudopotential is critical for the description short-range interactions in DFT~\cite{stoller_impact_2016, olsson2016ab}. 

Differences in the repulsive pair interactions have been demonstrated to have significant impact on the characteristics of displacement damage~\cite{terentyev2006effect,byggmastar_effects_2018,malerba_molecular_2006,stoller2016impact,de2021modelling,becquart_modelling_2021}. In tungsten it has been shown that ''harder'' potentials produce denser collision cascades, which in turn decreases the number of Frenkel pairs produced~\cite{sand_non-equilibrium_2016,de2021modelling}. Similar findings have been reported in iron, where however it was found that the effects were less significant in overlapping cascades once a sufficient dose was reached~\cite{byggmastar_effects_2018}. 

The development (and use) of machine-learning interatomic potentials (MLIP) for MD simulations has become commonplace in the past decade. This is due to their superior accuracy compared to classical analytical potentials. Furthermore, significant progress has been made in the computational efficiency of MLIPs, which has enabled their use in large-scale simulations. Most MLIPs that have been developed are, however, made with near-equilibrium simulations in mind. In extreme environment, such as those present in radiation damage simulations, the proper description of far-from-equilibrium phenomena becomes critical, usually requiring additional consideration by the developer of the MLIP. In recent years, MLIPs have been used in several studies of radiation damage~\cite{byggmastar2019machine,koskenniemi2023efficient,wei2023effects,liu2023large,niu2023machine,wang2022machine,wang2019deep} and this will no doubt increase in the future. Therefore, it is important to examine systematically how reliable  MLIPs are for radiation damage simulations.

The main assumption of this work is that the use of a different pseudopotential in the creation of training data for a MLIP has an impact on the predictions of the final model. This is a reasonable assumption especially in the case of radiation damage simulations, where short-range interactions are highly relevant. In this work, we investigate in detail the effects that the choice of the pseudopotential in DFT used in the generation of the training database of a MLIP can have on radiation damage simulations. We do this comparison in nickel, as radiation damage in nickel has been studied extensively using computational methods in the past. Furthermore, we can use as a starting point an MLIP from our previous work, making the re-computation of the training database and the training of the model straightforward. We recomputed the nickel DFT database using a ''harder'' semicore pseudopotential and retrained the model. Thus we get a suitable model to make 
detailed comparisons with, and -- as an added benefit -- we get a new model with improved short-range interactions. We use the models to study threshold damage, primary radiation damage and massively overlapping cascades. Additionally, we make direct comparisons of the overlapping cascades to experiments via Rutherford Backscattering Spectroscopy in channeling mode spectra (RBS/c). Furthermore, we test and discuss the feasibility of approaches to modify the repulsive pair interactions after the initial training of the models.  


\section{Methods}
\label{sec:methods}

\subsection{Tabulated Gaussian approximation potentials}\label{sec:met:tabgap}

The machine-learned potentials in this work are so called tabulated Gaussian approximation potentials (tabGAP)~\cite{byggmastar_simple_2022}. The tabGAP formalism is a low-dimensional tabulated version of a Gaussian approximation potential (GAP)~\cite{bartok_gaussian_2010}. The low-dimensional descriptors and tabulation allows for orders of magnitude speed up compared to the GAP potential, allowing for simulations of large system sizes. In this work we have taken the Ni tabGAP potential we developed previously~\cite{fellman2024fast} and recomputed its entire training and testing databases using the semicore pseudopotential of the projector augmented wave (PAW) semicore pseudopotential \texttt{Ni\_pv}. The previous potential in~\cite{fellman2024fast} used the standard \texttt{Ni} pseudopotential. Thus from this point on, we will refer to the previous potential from Ref.~\cite{fellman2024fast} as the Ni tabGAP and the new potential as the Ni\texttt{\_pv} tabGAP (or as the semicore potential) after the different pseudopotentials used in the training databases. Otherwise, the two different potentials are made in the exact same manner. We will now give a brief description of the tabGAP potentials. for a more detailed description we refer to Refs.~\cite{fellman2024fast,byggmastar2022multiscale}. 

The total energy for the to-be-tabulated GAP is
\begin{equation}
\begin{split}
E_\mathrm{3b+eam} = E_{\mathrm{rep.}}  &+  \sum_{i<j}^N \delta^2_\mathrm{2b} \sum_s^{M_\mathrm{2b}} \alpha_{s} K_\mathrm{se} (r_{ij}, r_s)  \\ 
& + \sum_{i, j < k}^N \delta^2_\mathrm{3b} \sum_s^{M_\mathrm{3b}} \alpha_{s} K_\mathrm{se} (\bm{q}_{ijk}, \bm{q}_s) \\ 
& + \sum_{i}^N \delta^2_\mathrm{eam} \sum_s^{M_\mathrm{eam}} \alpha_{s} K_\mathrm{se} (\rho_i, \rho_s) ,
\end{split}
\label{eq:tabgap}    
\end{equation}
where the sum over $N$ refers to the sum over local descriptor environments and $M$ refers to a selected sparsified subset of descriptor environments from the training structures. The two-body term consists of prefactors $\delta^2_\mathrm{2b}$ and regression coefficients $\alpha_{s}$. Furthermore, it uses a squared exponential kernel $K_\mathrm{se}$ and a two-body descriptor that corresponds simply to the distance between two atoms, as well as a three-body cluster descriptor (defined as a three-valued permutation-invariant vector~\cite{bartok_gaussian_2015}) and an embedded atom method (EAM) density descriptor. The EAM density is a scalar pairwise summed radial function, as in standard EAM potentials~\cite{byggmastar_simple_2022,daw_embedded-atom_1984,finnis_simple_1984}.

Additionally, the potentials include an external purely repulsive potential. The repulsive part is included in order to more accurately describe short-range interactions. The repulsive part in our case is a Ziegler-Biersack-Littmark-type (ZBL) repulsive potential:
\begin{equation}
E_{\mathrm{rep.}} = \sum_{i<j}^{N} \frac{1}{4 \pi \epsilon_0} \frac{Z_i Z_j e^2}{r_{ij}} \phi (r_{ij}/a) f_{\mathrm{cut}}(r_{ij}),
\end{equation}
where
\begin{equation}
    a = \frac{0.46848}{Z_i^{0.23} + Z_j^{0.23}}.
\end{equation}

The screening function $\phi (r_{ij}/a)$ was refitted to repulsive dimer data from all-electron DFT calculations (DMol)~\cite{Nor96c,nordlund2025repulsive}. Additionally, the screened Coulomb potential is multiplied by a cutoff function to force it to zero well below the nearest-neighbour distance of the material, to avoid interfering with the near-equilibrium interactions described by the machine-learned part. The range of the cutoff was chosen to be 1.0--2.2 Å to ensure smooth transition between the all-electron calculations and short-range DFT data that was part of the training data. Recently, the repulsive pair interactions have been calculated with even greater accuracy than the all-electron DMol calculations used in the fitting of the external repulsive potential~\cite{nordlund2025repulsive}. In that work it was found that the older DMol calculations are in general accurate to 2\% compared to the new second-order Møller-Plesset perturbation theory calculation.

The tabGAP is trained as a GAP above, but after the initial training the energy contributions of the different terms of the potential are tabulated onto low-dimensional grids and evaluated using cubic-spline interpolation. The repulsive and two-body terms are tabulated into a one-dimensional grid, the three-body term into a three-dimensional grid and the EAM term into two one-dimensional grids. Further details on the tabulation process can be found in Ref.~\citenum{byggmastar2022multiscale}. 

The hyperparameters of the GAP were kept the same as for the original potential and are given in the supplementary materials and in Ref.~\cite{fellman2024fast}. Furthermore, the training and testing databases were identical between the two tabGAPs, with the exception of a handful of structures that did not converge with the new DFT parameters and were omitted. The database includes short-range structures relevant for radiation damage simulations. A more detailed description of the training database is given in the supplementary materials and in Ref.~\cite{fellman2024fast}. Additionally, we kept the same test-train split between the models. All this was done to create a new tabGAP version that was as close to the original model as possible in all other respects than the choice of pseudopotential used in the DFT calculations of the training database.

\subsection{Density functional theory calculation details}\label{sec:dft-calc}

All the DFT calculations in this work were performed using the \textsc{VASP} DFT code~\cite{kresse_ab_1993,kresse_ab_1994,kresse_efficiency_1996,kresse_efficient_1996}. The calculations used the PBE GGA exchange-correlation function~\cite{perdew_generalized_1996} and projector augmented wave (PAW) pseudopotentials \texttt{Ni} and the semicore \texttt{Ni\_pv} potential. The plane-wave expansion energy cutoff was set to 500 eV and 650 eV respectively. The increased cutoff for the semicore potential was chosen as a consequence of convergence testing. K-points were defined using $\Gamma$-centered Monkhorst–Pack grids~\cite{monkhorst_special_1976} with maximum k-spacing of 0.15 Å$^{-1}$. Additionally, first order Methfessel-Paxton smearing of 0.1 eV was applied~\cite{methfessel_high-precision_1989}. The calculations were performed with spin-polarization with initial ferromagnetic order.

\subsection{Threshold displacement energy calculations}

The threshold displacement energy calculations were performed in the same manner as for the original Ni tabGAP potential~\cite{fellman2024fast}. All molecular dynamics simulations used the LAMMPS simulation package.~\cite{LAMMPS}. The threshold displacement energies were determined by sampling 800 random spherically uniformly distributed lattice directions. A series of simulations were performed for each direction, where an atom is given higher and higher kinetic energy (2 eV increments) in the specific direction until a stable defect is formed. During this simulation, border cooling was applied ($NVT$) with a temperature of 10 K, while the rest of the system was kept in the $NVE$ ensemble. The simulation cell was chosen to be ($12 \times 13 \times 14$) units cells which corresponds to 8736 atoms.
A non-cubic simulation cell size was specifically chosen to reduce the possibility that replacement collision sequences along low-index crystal directions interact with themselves across the periodic boundaries.

\subsection{Cascade simulations}

Both primary cascade damage and massively overlapping cascade simulations \cite{Nord01,granberg_mechanism_2016} were performed with the different potentials. For all cascade simulations the initial samples were relaxed to 300 K before irradiation. The individual cascade events are initiated by choosing a primary knock-on atom (PKA) close to the middle of the simulation cell and giving it a corresponding kinetic energy in a random direction sampled from a uniform spherical distribution. During the cascade event, electronic stopping was applied on atoms with a kinetic energy above 10 eV, implemented as a frictional term using pre-calculated stopping data. The electronic stopping data were computed using the SRIM electronic stopping model~\cite{ziegler_srim_2010}. During cascade simulations, the boundaries of the system were connected to a thermostat ($NVT$) to simulate the dissipation of temperature to the surrounding material. Self-interaction over the periodic boundary was limited by halting and restarting the cascade event if an atom with over 10 eV energy crossed the simulation cell border. For the simulation of primary damage we used a simulation time of 50 ps during which the simulation cell had ample time to cool down back to the initial temperature of 300 K. The system size was 2\,048\,000 atoms for 10 keV cascades and 256\,000 atoms for PKA energies of 5 keV and below. In the case of massively overlapping cascades, we used a PKA energy of 5 keV and system size of 256\,000 atoms. For each individual overlapping simulation, the simulation cell was randomly shifted, atoms that crossed the boundaries were returned back to the other side according to the periodic boundaries \cite{Nord01} and a cascade event was initiated in the centre of the cell. This was done in order to uniformly distribute the recoils within the systems. The cascade event was simulated for 20 ps, with an additional 10 ps of relaxation ($NPT$) back to the initial 300 K temperature and zero overall pressure. The final configurations were analysed using the Wigner-Seitz (WS) method~\cite{nordlund_defect_1998} to determine the number of Frenkel pairs produced from the cascade events. Furthermore, DXA~\cite{0965-0393-18-8-085001} analysis was performed, implemented in OVITO~\cite{stukowski2012automated}, in order to find the types and lengths of the dislocations.

\subsection{Rutherford backscattering calculations}

\begin{figure}
    \centering
    \includegraphics[width=0.99\linewidth]{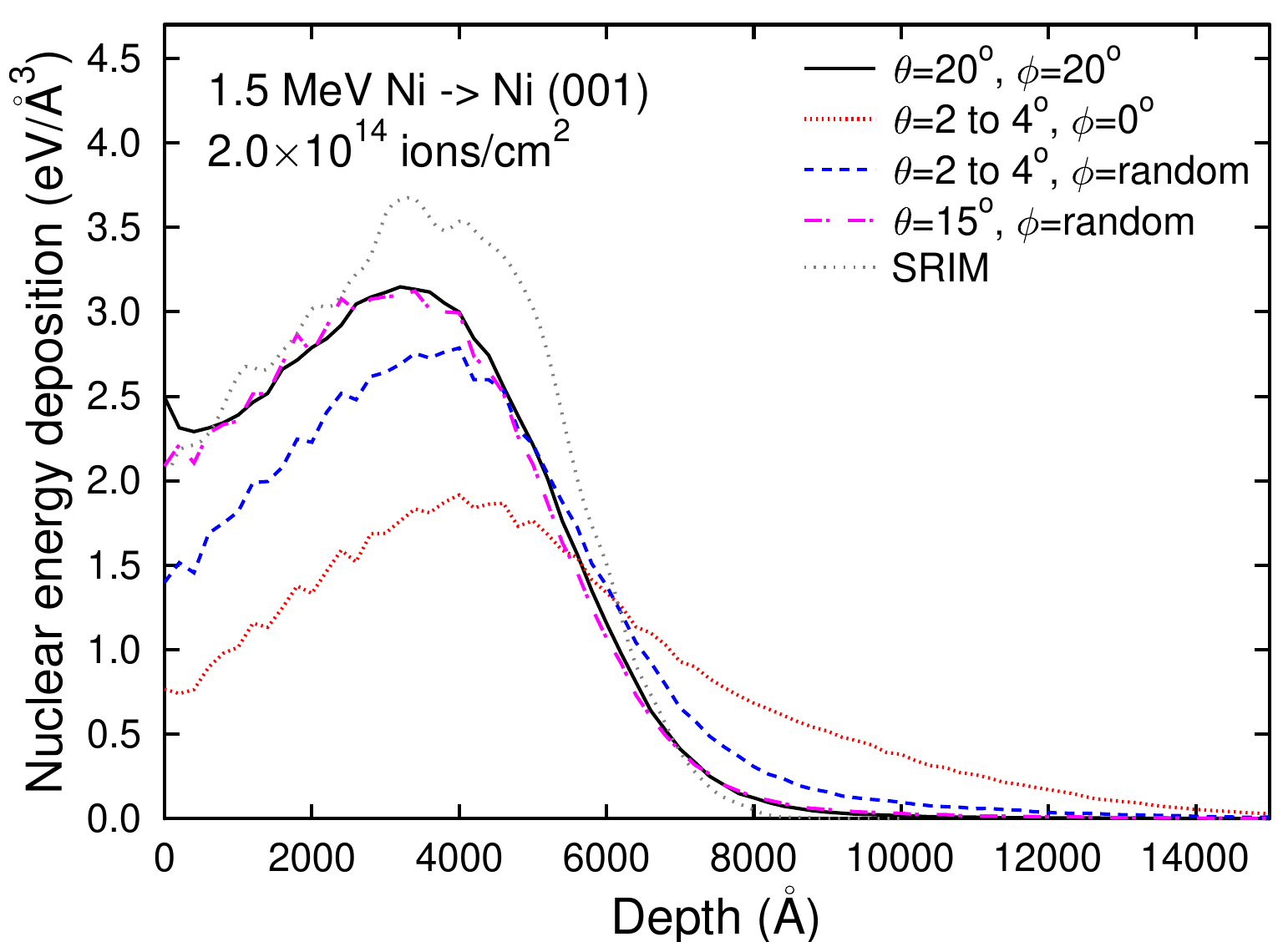}
    \caption{Calculated nuclear energy deposition profiles.}
    \label{fig:depens}
\end{figure}

As a comparison with experimental results we carried out RBS/c calculations using the RBSADEC code~\cite{zhang2016simulation,jin2020new}. The simulation parameters and procedures followed closely those detailed by Levo \etal~\cite{levo2021temperature}, which in turn is based on previous work in Ref.~\cite{zhang2017radiation}. In the calculations 3.5 MeV He$^+$ ions were used as the backscattering ions. The temperature during the calculation was set to 300 K and the calculation used 4096 as the number of channels, each corresponding to 1 keV. The simulation samples were created by stacking simulation boxes from the overlapping cascade simulations based on a given nuclear energy deposition depth profile. We merged 62 irradiated simulation cells, resulting in a final merged cell nearly 900 nm long. We considered two different nuclear energy deposition profiles. First, we used the same nuclear deposition profile that was used in the work by Levo \etal. However, this profile used a fluence that was half of the experimental results and the dose was estimated using NRT-dpa~\cite{nordlund_primary_2018} with the assumption of 40 eV TDE. In addition to the profile taken from Levo \etal, we recalculated the nuclear energy deposition profile to more closely match the experimental conditions. To calculate the nuclear energy deposition by the 1.5 MeV Ni ions, we used MDRANGE~\cite{Nor94b} with the new NLH interatomic potentials~\cite{nordlund2025repulsive}. Since in the experiments the 1.5 MeV Ni ion beam was tilted a few degrees off normal~\cite{Zha15b}, we modelled the nuclear energy deposition with a tilt angle randomly chosen between 2 and 4 degrees off normal, with no twist. This gives an energy deposition that is considerably lower than in a clearly non-channeling direction. The MDRANGE code gives the energy to primary recoils, which, however, still undergo electronic stopping themselves. Full cascade SRIM2013 simulations~\cite{SRIM-2013} showed that the electronic energy deposition of the recoils from 1.5 MeV Ni ions is about 11.5\%. Hence the MDRANGE nuclear energy deposition curve was still multiplied by 0.885. Fig.~\ref{fig:depens} shows the new calculations of the nuclear energy deposition profiles used in this work. The $\theta = \mathrm{2~to~4}\degree$ and $\phi = 0\degree$ case was used for the sampling.

\subsection{Modifications to the repulsive pair interactions}\label{sec:met:mod}

At the end of the results section (\ref{sec:res:mod}), we discuss the possibility and feasibility of approaches for the modification (post-training) of repulsive interactions of the MLIP. These modifications are detailed in this section. We considered two approaches. 

The first being akin to the process commonly employed with classical potentials called ''hardening'' or ''stiffening'' that has been widely used to include repulsive pair interactions~\cite{terentyev2006effect,bjorkas_modelling_2009,sand_non-equilibrium_2016,byggmastar_effects_2018,stoller2016impact}. In this case we take the external repulsive potential used in the training of the tabGAP potential and refit the two-body repulsion. The ''refitting'' is done by replacing the two-body grid values at short-distances with the those computed with the purely repulsive external potential, which has been presented in the section~\ref{sec:met:tabgap}. These are then smoothly joined to the two-body grid values close to the equilibrium distance with a switching function. The switching function is the cutoff function commonly used in classical potentials:

\begin{equation}
    f_c(r_{ij}) = \begin{cases}
    1, & r_{ij} \leq r_c - r_d \\
    \frac{1}{2} \bigg( 1 - \sin \frac{\pi (r_{ij} -r_c)}{2 r_d} \bigg), & |r_c - r_d| \leq r_d \\
    0, & r_{ij} \geq r_c + r_d
    \end{cases},
\end{equation}
where we have the parameters $r_c$ (center of cutoff region) and $r_d$ (width of cutoff region). Thus we have two parameters that we need to define during the refitting. The potential of this kind is denoted in section~\ref{sec:res:mod} as Ni$_{\mathrm{reppot}}$. 

The other approach is to ''merge'' or smoothly transition between the two different MLIPs at shorter distances. This approach has been employed in MLIP models in the literature~\cite{wang2022machine}. In our case we only transition between the two-body grid values of two different tabGAPs. In the potential denoted Ni$_{\mathrm{merged}}$ the Ni\texttt{\_pv} tabGAP's two-body grid points define the short-range interactions and are smoothly transitioned to the Ni tabGAP when approaching near-equilibrium interatomic distances. In the potential denoted Ni\texttt{\_pvsoft} the Ni tabGAP defines the short-range two-body grid points and is smoothly transitioned to the  Ni\texttt{\_pv} potential towards the equilibrium distances. In all cases the parameters of the joining (i.e. $r_c$ and $r_d$), where defined to give reasonable dimer curves and quasi-static drag (QSD) curves. Quasi-static drag is a simulation where a single atom is moved along a given direction while keeping all other atoms stationary. These parameters and curves are given in the supplementary materials. 

\section{Results and discussion}

The fundamental question we want to answer in this work is: what changes in the radiation damage simulation results if we train MLIPs with different pseudopotentials? In order to answer this, we will present results and discuss differences starting from fundamental materials properties, then threshold displacement energies and primary radiation damage, and finally large-scale overlapping cascade simulations. In these results we also compare against two classical EAM potentials, the Stoller \etal potential~\cite{stoller2016impact} and the Bonny \etal potential~\cite{bonny2013interatomic}. We choose Stoller \etal as its method of connecting the equilibrium and repulsive parts is well motivated. Additionally, we compare with the Bonny \etal potential as it has been used in previous studies of radiation damage simulations in Ni~\cite{Zha15b,zarkadoula2017effects,levo2021temperature}.  

\subsection{Dimer curves}

\begin{figure}
    \centering
    \includegraphics[width=0.99\linewidth]{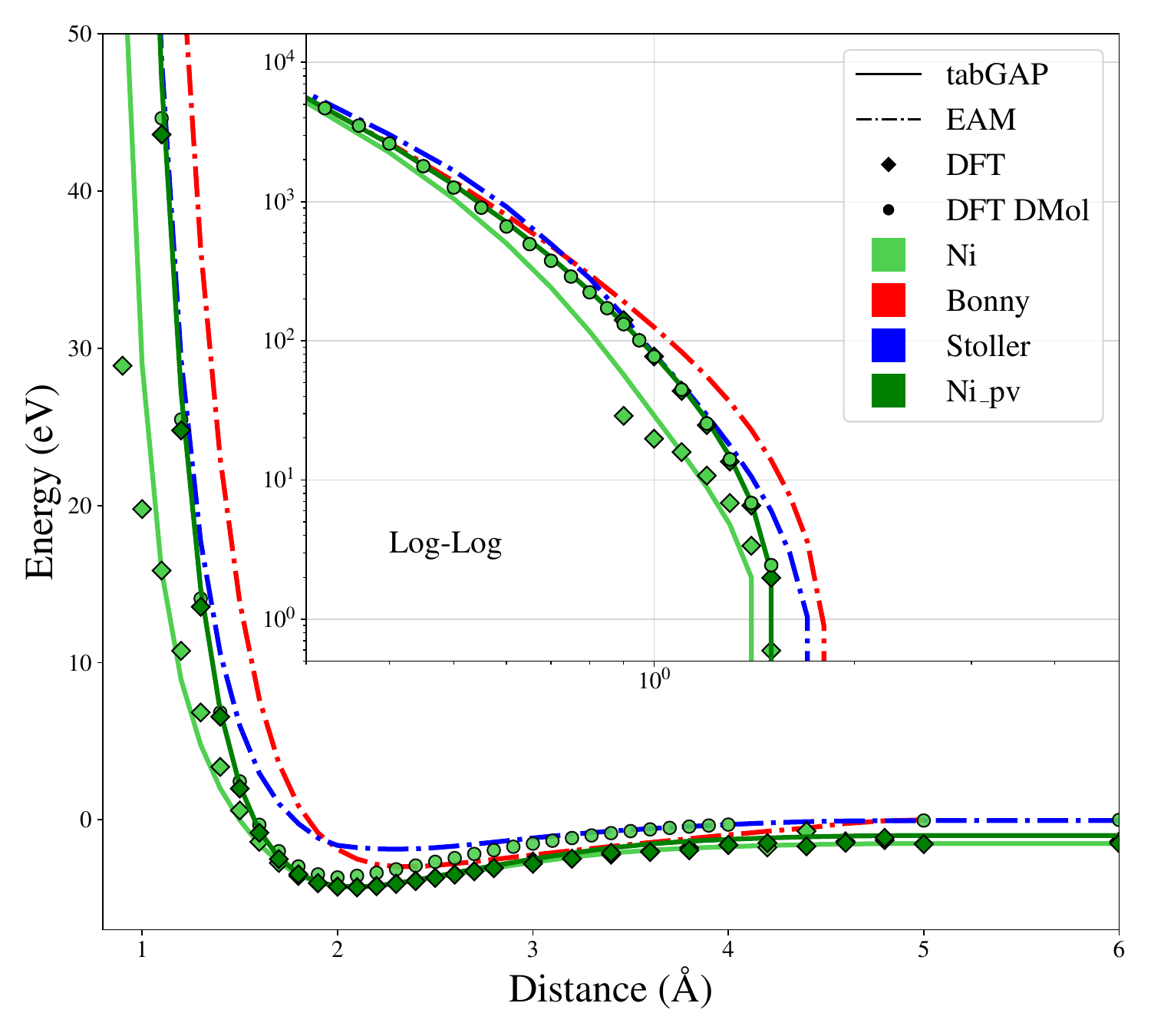}
    \caption{Dimer curves calculated with the developed Ni and Ni\texttt{\_pv} tabGAP potentials as well as EAM potentials compared with DFT and all electron DFT ("DFT DMol") used in external repulsive potential.}
    \label{fig:dimer}
\end{figure}

The most basic test of a potential in regards to its applicability to simulate radiation damage is to verify that the repulsive interactions of the potential are properly described. Fig.~\ref{fig:dimer} shows the dimer curves calculated with the Ni potentials compared with both DFT and all-electron DFT used in the external repulsive potentials of the tabGAP potentials~\cite{Nor96c}. We can immediately see that both the tabGAP potentials follow their respective DFT data points with good accuracy, which is to be expected. However, when we plot the dimer curves in a log-log scale, we emphasize the differences between the two tabGAP potentials. We see that the Ni\texttt{\_pv} DFT data smoothly transitions towards the all electron DFT data. Furthermore, the Ni tabGAP can be seen to be somewhat ''softer'' in the range of 0.5--1.5 Å. The differences in the repulsive interactions of interatomic potentials has been discussed in great detail in the literature and the general conclusion is that the repulsive interactions have a strong impact on the characteristics of displacement damage~\cite{terentyev2006effect,byggmastar_effects_2018,malerba_molecular_2006,stoller2016impact,de2021modelling,becquart_modelling_2021}. 

There are several implications and lessons to draw from Fig.~\ref{fig:dimer}. First, we need to emphasize that since the external repulsive potentials in the two tabGAP potentials are identical, the difference we see is from the machine-learned transition towards the external repulsive potential. Note that the cutoff of the external repulsive potential was between 1.0--2.2 Å. This implies that during training the ML part of the Ni tabGAP has ''softened'' the overall potential in the region between  0.5--1.5 Å. As the training data includes short-range structures with atoms within this region, this is unsurprising. It is of course possible to curate the short-range structures included in the training dataset and/or optimize the regularization parameters of those structures and to finetune the cutoff of the external repulsive potential to improve this transition. However, this can become quite tedious and in our experience the inclusion of the short-range structures improves the potential stability of the short-range interactions and inhibits the crazy extrapolations that can occur with MLIPs if they are used for interatomic distances not at all included in the training set. Second, if only looking at DFT data used in the training of the potential, we might think that the potential transitions towards the ZBL properly. The broad lesson is that just because a ML potential has a repulsive potential (ZBL or similar), does not automatically guarantee that the repulsive interactions are properly represented. Most likely the creator of the potential has to have included short-range interaction in the training data of the potential, which adds its own challenges in the training of the model. However, while the differences in dimer repulsion is nothing new and its implications have been discussed in great detail~\cite{terentyev2006effect,byggmastar_effects_2018,malerba_molecular_2006,stoller2016impact,de2021modelling,becquart_modelling_2021}, it does not tell the whole story, which we shall present in the subsequent sections. 

\subsection{Basic properties}\label{sec:basic}

%
%
A standard check of a ML potential is to calculate the root-mean-squared (RMS) error of the potential compared to the testing structures. Table~\ref{tab:Test_errors} gives the RMS errors of the energies and forces of the two tabGAP potentials with identical test set. We further divide these errors into crystalline and liquid structures as the liquid structures are relevant to radiation damage simulations and are more difficult to represent. The short-range structures are not included in these RMS errors as as they contain by design very strong forces and high energies that are not, and need not, be trained accurately. What is striking is the fact that the RMS energy errors of the Ni\texttt{\_pv} potential are a higher than that of the Ni tabGAP, while there is only small differences in the liquid force errors between the potentials. Most likely this is due to the fact that suddenly the short-range structures have become higher in energy which impacts the training. Regardless, the RMS errors are a quite limited form of validation and blindly relying on them as a benchmark can be misleading. This is especially the case with potentials made for extreme conditions (such as radiation damage), where we want to include extreme far-from-equilibrium structures both in the training and testing data.

\begin{table}[ht!]
 \caption{Energy and force RMS errors of the testing data for crystalline and liquid structures separately.}
 \label{tab:Test_errors}
 \begin{threeparttable}
  \begin{tabular}{|lcc|cc|}
   \toprule
   &   Energy (meV/atom) & &  Force (eV/Å) &   \\
   &  $E_{\mathrm{crystalline}}$ & $E_{\mathrm{liquid}}$  & $F_{\mathrm{crystalline}}$ & $F_{\mathrm{liquid}}$ \\
   \bottomrule
    Ni&0.6 &4.7&0.02&0.07\\
    Ni\texttt{\_pv}& 2.3 & 6.5 & 0.02 & 0.08 \\
   \bottomrule
  \end{tabular}
 \end{threeparttable}
\end{table}

%
%
Table~\ref{tab:bulk} shows basic material properties computed with the tabGAPs. Here we mainly want to highlight the differences between the two different tabGAP potentials, or more precisely in this case, the lack thereof. The elastic constants and surface energies are very similar in the potentials. There are some slight differences in the formation energies of point defects. Furthermore, the melting point, simulated with the two-phase method~\cite{Mor94} with a temperature increment of 20 K, is also very similar. 

\begin{table*}
 \caption{Calculated material properties compared with DFT and experimental results. $a$: Lattice constant (Å),  $E_\mathrm{mig}^\mathrm{vac}$: Vacancy migration energy (eV), $C_{ij}$: Elastic constants (GPa), $\gamma_{(hjk)}$: Surface energies(mJ$/\mathrm{m}^2$), $E_\mathrm{f}$: Formation energy of a single vacancy and interstitials (eV), $T_{\mathrm{melt}}$: Melting temperature (K) }
 \label{tab:bulk}
 \begin{threeparttable}
  \begin{tabular}{llcccccr}
   \toprule
   &Property& tabGAP\tnote{a}~~~  & tabGAP\texttt{\_pv}~~~ & Stoller\tnote{b}~~~ & Bonny\tnote{c}~~~~ & DFT~~~  & Expt.\\
   \midrule
    &$a$ (Å)  & 3.518 & 3.517 & 3.52 & 3.522 & 3.519\tnote{a} & 3.524\tnote{d}\\
    &$E_\mathrm{mig}^\mathrm{vac}$ (eV) & 1.06 & 1.04 & 1.18 & 1.11 & 1.12\tnote{e} &  1.04\tnote{f}\\ 
    &$C_{11}$ (GPa) & 281 & 289 & 241 & 247 &273\tnote{a} & 247\tnote{d} \\ 
    &$C_{12}$ (GPa) & 156 & 152 & 150 & 144 &155\tnote{a} & 153\tnote{d} \\ 
    &$C_{44}$ (GPa) & 125 & 125 & 127 & 107  &131\tnote{a} & 122\tnote{d}\\ 
    &$\gamma_{(100)}$ (mJ$/\mathrm{m}^2$)& 2208 & 2248 & 1936 & 1832 & 2210\tnote{g} & 1940\tnote{h,*} \\ 
    &$\gamma_{(110)}$ (mJ$/\mathrm{m}^2$)& 2272 & 2291 & 2087 & 1940 & 2290\tnote{g} & - - \\  
    &$\gamma_{(111)}$ (mJ$/\mathrm{m}^2$) & 1910 & 1918 & 1759 & 1764 & 1920\tnote{g} & - - \\ 
    &$E^\mathrm{vac}_\mathrm{f}$ (eV)& 1.47 & 1.44 & 1.57 & 1.39 & 1.49\tnote{a} & 1.79 $\pm$ 0.05\tnote{d} \\
    &$E^\mathrm{100d}_\mathrm{f}$ (eV)& 4.13 & 4.20 & 3.95  & 5.85 & 4.07\tnote{a} & -  \\
    &$E^\mathrm{octa}_\mathrm{f}$ (eV)& 4.35 & 4.42 & 4.29 & 5.94 & 4.26\tnote{a}  & -\\
    &$E^\mathrm{tetra}_\mathrm{f}$ (eV)& 4.73 & 4.83 & 4.43 & 7.0 & 4.67\tnote{a} & - \\
    &$T_{\mathrm{melt}}$ (K) & 1690 & 1725 & 1770 & 1510 & -  &1726\tnote{i}\\ 
   \bottomrule
  \end{tabular}
  \begin{tablenotes}[flushleft]
   \item[a] \cite{fellman2024fast}
   \item[b] \cite{stoller2016impact}
   \item[c] \cite{bonny2013interatomic}
   \item[d] \cite{ma_nonuniversal_2021}
   \item[e] \cite{zuo_performance_2020}
   \item[f] \cite{LandoltBornstein1991:sm_lbs_978-3-540-48128-7_63}
   \item[g] \cite{tran2016surface} 
   \item[h] \cite{kumikov1983measurement}   
   \item[i] \cite{haynes_crc_2015}
   \footnotesize
   \item[*] Not specific direction. 
  \end{tablenotes}
 \end{threeparttable}
\end{table*}

\begin{figure}
    \centering
    \includegraphics[width=0.8\linewidth]{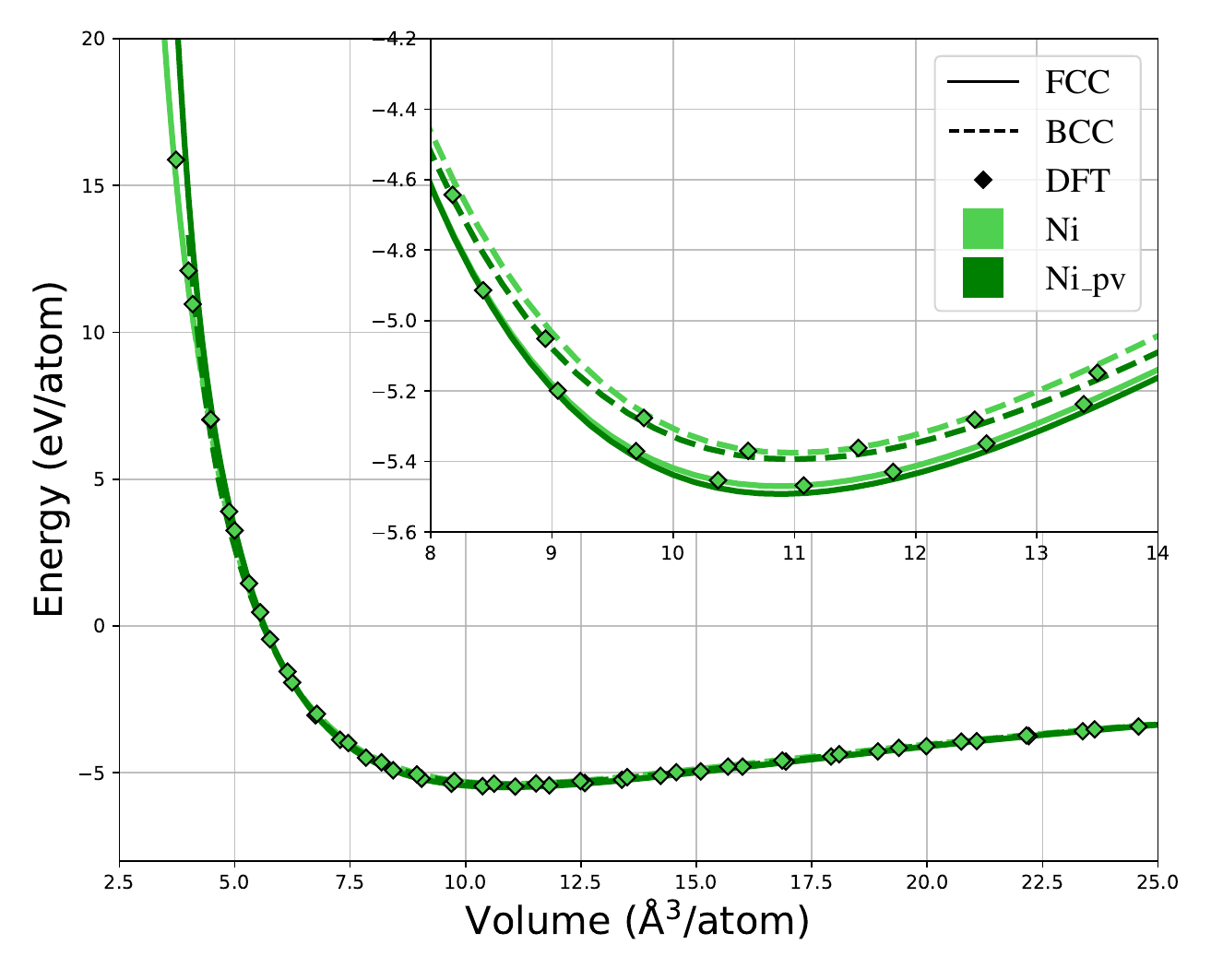}
    \caption{Energy vs. volume of two different Ni tabGAP potentials compared with DFT calculations.}
    \label{fig:Eng-Vol}
\end{figure}

%
%
Another, typical validation is to compute the volume-energy curves of the material and compare them with DFT calculations. Fig.~\ref{fig:Eng-Vol} shows the computed volume energy curves in both FCC and BCC lattice structure for the potentials. The curves around the equilibrium distance are pretty much identical. Only at energies of over 20 eV/atom do we start seeing differences similar to those reported for the dimer curves, but they are less noticeable. Volume energy curves should be provided when demonstrating a MLIPs suitability for radiation damage simulations as they demonstrate that the potential is stable at high densities. Especially, for MLIPs designed to be used in radiation damage simulations, the transition to the external repulsive potential is crucial. The equilibrium part is often very easy for the MLIPs to get correct as there is usually lots of data in those volume ranges. In practice, the addition of high-energy close-range structures makes it more tricky to get the whole range of volumes to behave appropriately and often over-fitting to close-range structures is noticeable in the volume energy curves. MLIPs can also fail catastrophically at high energies, due to high degree of extrapolation, and care must be taken to ensure proper behavior. However, in this case it is still quite difficult to distinguish between the two models purely on basis of the volume energy curves.   

\begin{figure}
    \centering
    \includegraphics[width=0.99\linewidth]{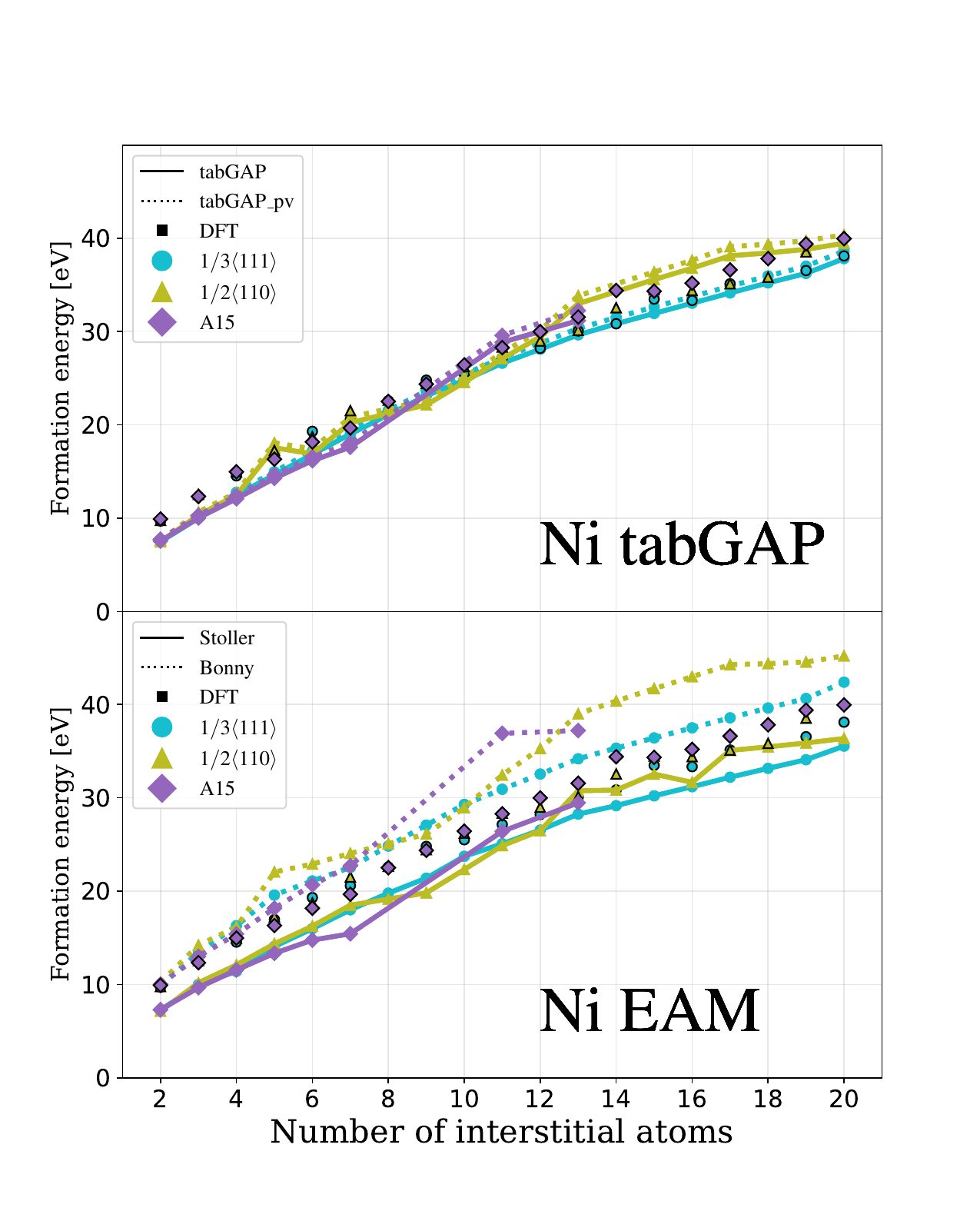}
    \caption{Defect formation energies of A15 clusters and interstitial dislocation loops compared with DFT calculations (lowest energy data points only) from Ref.~\cite{goryaeva2023compact}. Markers and line styles indicate the method used and the colors indicate the defect type.}
    \label{fig:def-form}
\end{figure}

%
%
Recently, studies of the A15 Frank-Kasper nano-phases has shed new light on the formation of dislocation loops in FCC metals~\cite{goryaeva2023compact,jourdan2024preferential}. Fig.~\ref{fig:def-form} shows the defect formation energies of A15 clusters and interstitial-type dislocation loops compared with DFT calculations from Ref.~\cite{goryaeva2023compact}. Here we followed closely the calculation of formation energies detailed in the publication, with system size of $13500 + N$, where $N$ is the number of interstitials. Conjugate gradient relaxation was performed with a 0.002 eV/Å force convergence criterion. We can conclude that the Ni\texttt{\_pv} potential is systematically slightly higher in the formation energies compared to the Ni tabGAP, which was also the case for the interstitial formation energies given in Table~\ref{tab:bulk} and is likely a direct consequence of the harder DFT pseudopotential. However, this difference is quite negligible. 

\begin{figure}
    \centering
    \includegraphics[width=0.99\linewidth]{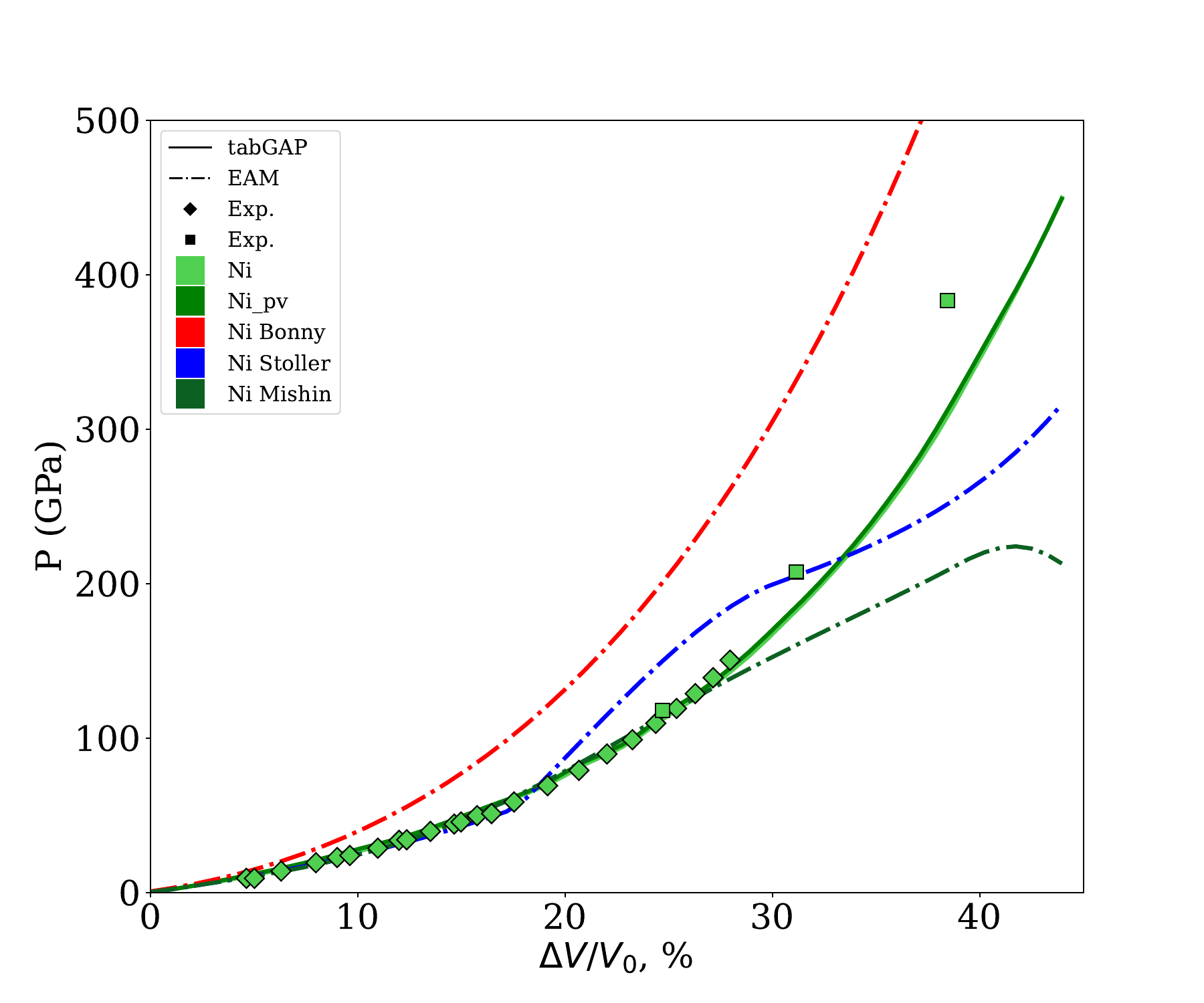}
    \caption{Equation of state calculated with various different interatomic potentials compared to experimental data. The different markers for experimental data for Ni are from different experiments~\cite{RICE19581,al1962shock}. Comparison is made to previously developed potentials: EAM~\cite{PhysRevB.59.3393,stoller2016impact,bonny2013interatomic}}
    \label{fig:eos}
\end{figure}

Furthermore, we can do validation of the developed potentials versus experiment by comparing the equation of states, as measured in experiment and simulated using the potentials. Fig.~\ref{fig:eos} shows the pressure-volume relation calculated with the different potentials compared with experimental data. Interestingly, there is not a significant difference between the tabGAP potentials even at high compressions. From the figure we see that the tabGAPs can reproduce very well the experimental results and there is quite large variations in the results from EAM potentials. In addition to the properties presented in this section, we have calculated the phonon dispersion curves and generalized stacking fault energy (GSFE) curves with the two tabGAP potentials. These can be found in the supplementary material, but the main observation is that there are no significant differences between the two tabGAP potentials. Based on this we conclude that in regards to basic materials properties the two tabGAP models are pretty much identical and differences between them are quite difficult to find. 

\subsection{Quasi-static drag}

\begin{figure*}
    \centering
    \includegraphics[width=0.32\linewidth]{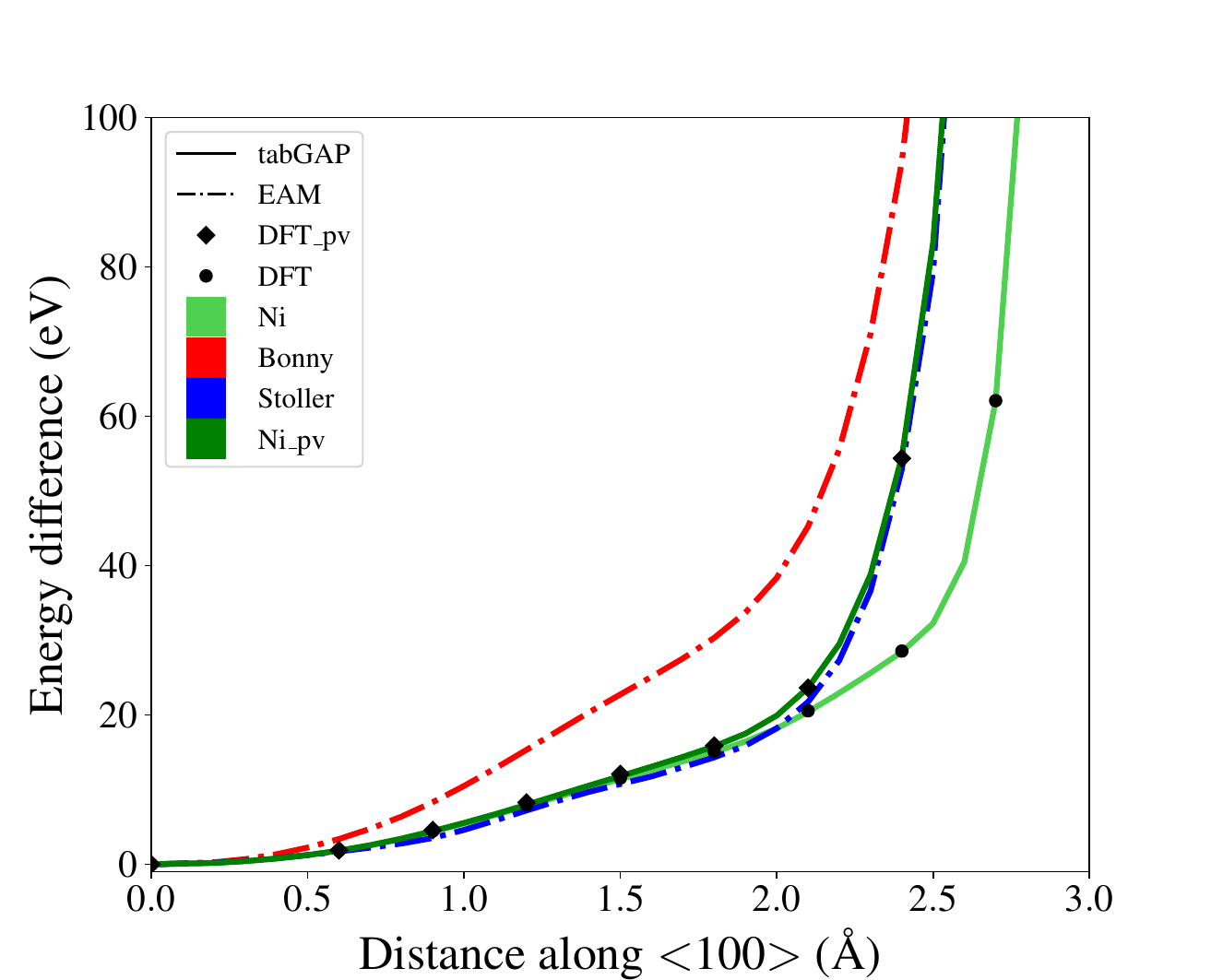}
    \includegraphics[width=0.32\linewidth]{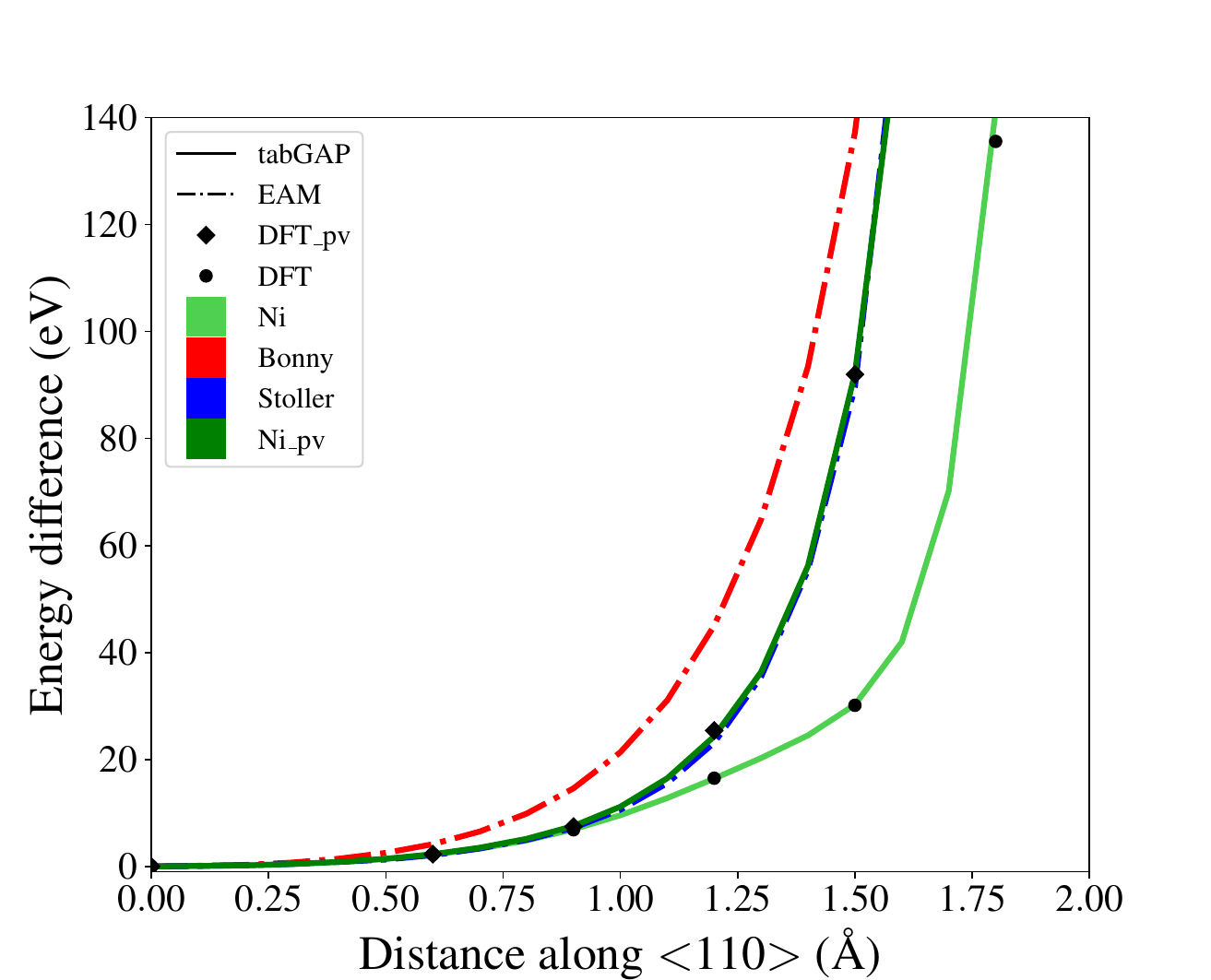}
    \includegraphics[width=0.32\linewidth]{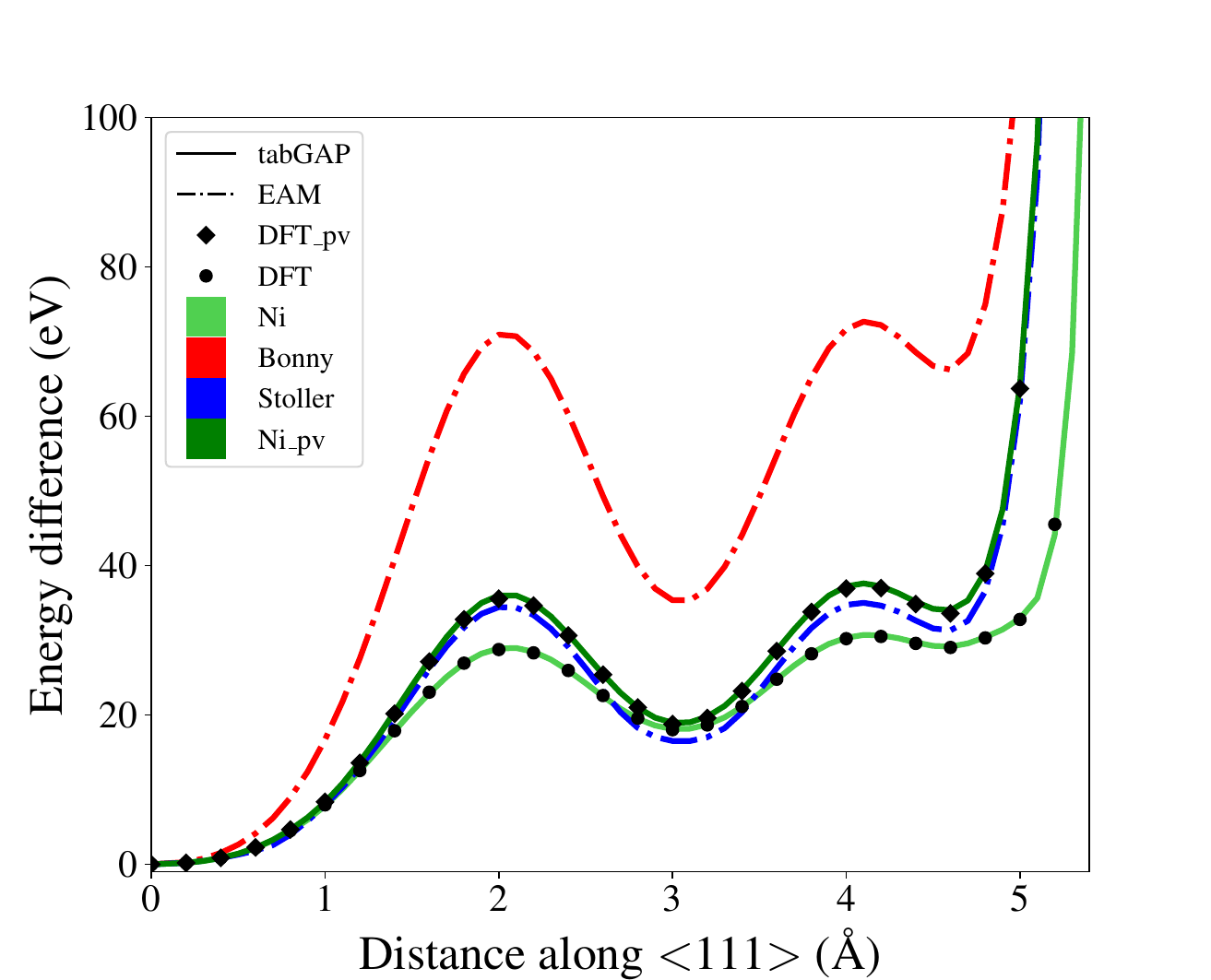}
    \caption{Total energy difference when displacing a single atom in different high symmetry directions while holding the other atoms stationary.}
    \label{fig:trajectory_Ni}
\end{figure*}

 In the past quasi-static drag simulations have been shown to correlate with the potentials ability to model displacement events~\cite{becquart_modelling_2021}. Fig.~\ref{fig:trajectory_Ni} shows the total energy difference from such trajectories in the high-symmetry directions compared with DFT. Looking at the figure we observe that the two tabGAP potentials follow very closely their respective DFT values. This is to be expected, as they include such structures in the training data. Additionally, we see that the two different pseudopotentials are reasonably similar until we get to the highly repulsive part. The figure also shows the remarkably similar values between the Ni\texttt{\_pv} and Stoller \etal potentials. Furthermore, as we saw that the Bonny potential is a clear outlier.

\subsection{Threshold displacement energies}\label{sec:tde}

\begin{figure*}[ht!]
    \centering
\begin{subfigure}[t]{0.49\textwidth}
\label{fig:tde:nitabgap}
    \includegraphics[width=\textwidth]{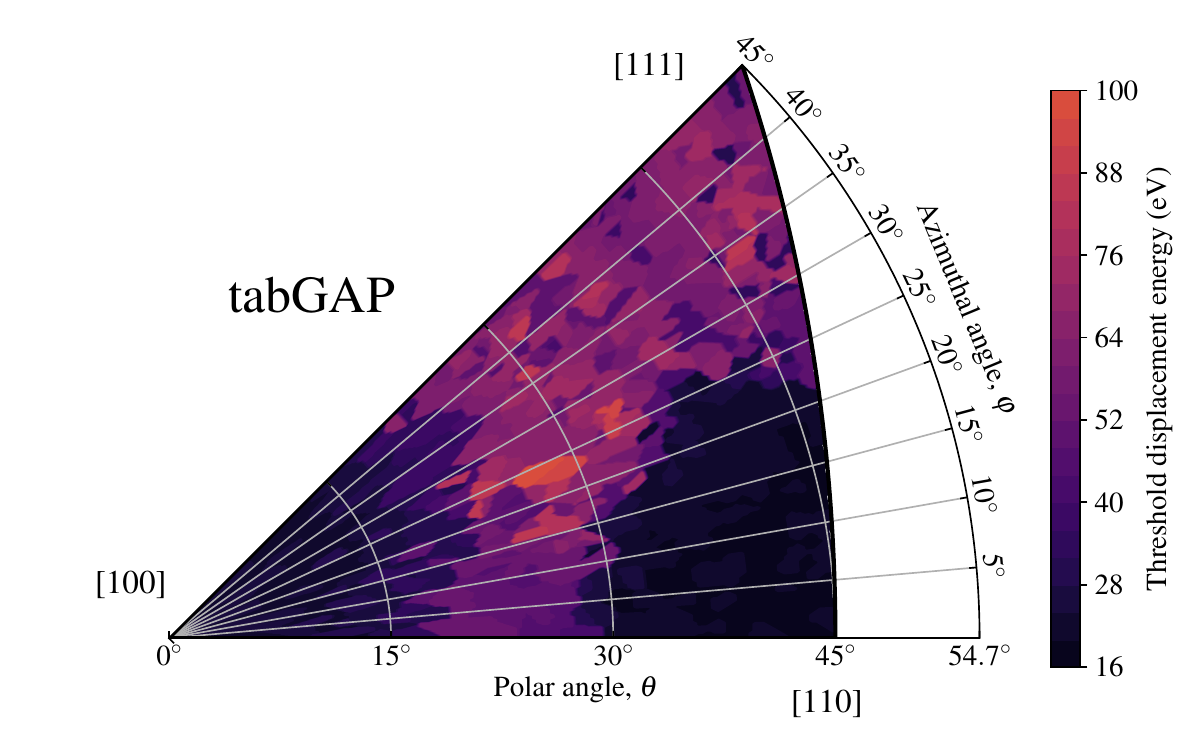}
      \caption{}
\end{subfigure}
\begin{subfigure}[t]{0.49\textwidth}
\label{fig:tde:nipvtabgap}
    \includegraphics[width=\textwidth]{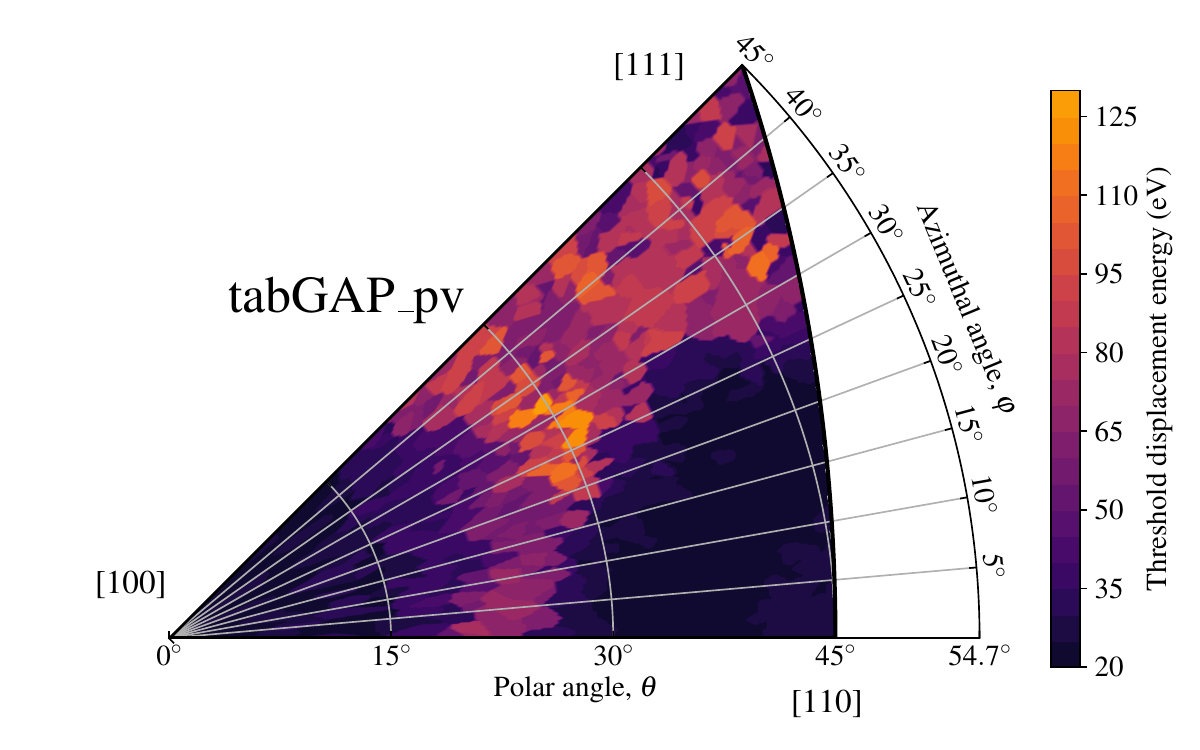}
    \caption{}
\end{subfigure}
\begin{subfigure}[t]{0.49\textwidth}
\label{fig:tde:stoller}
    \includegraphics[width=\textwidth]{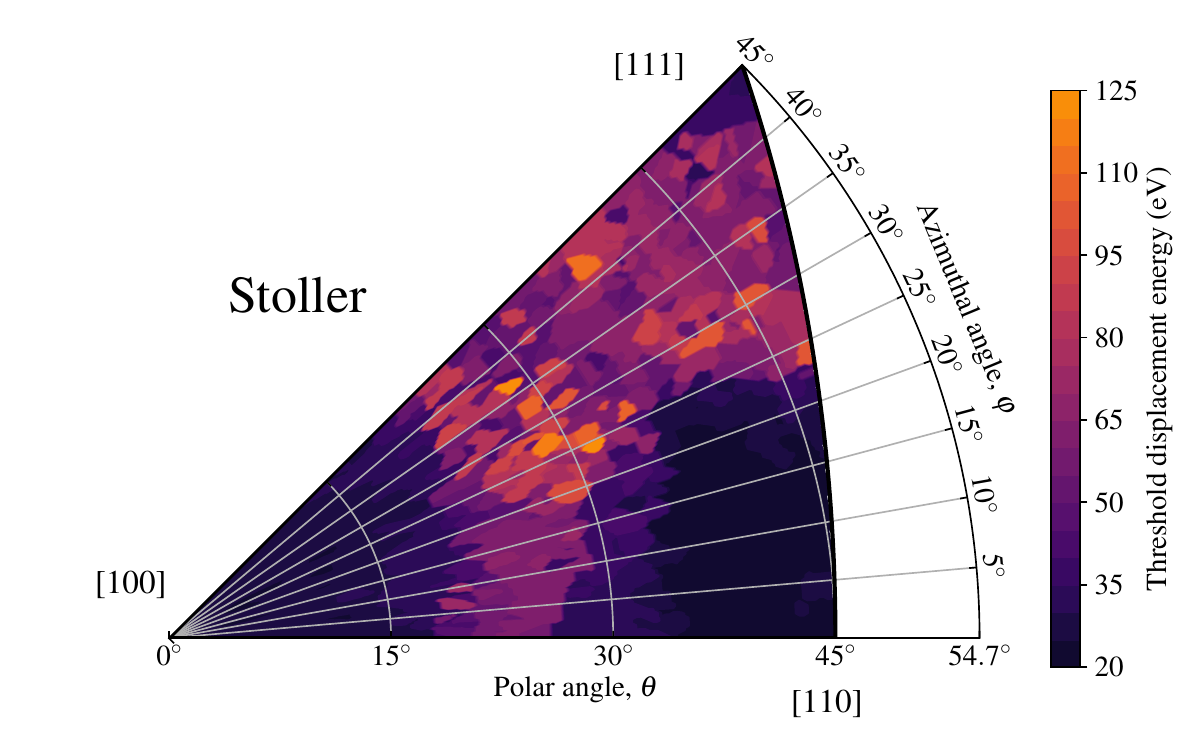}
    \caption{}
\end{subfigure}
\begin{subfigure}[t]{0.49\textwidth}
\label{fig:tde:bonny}
    \includegraphics[width=\textwidth]{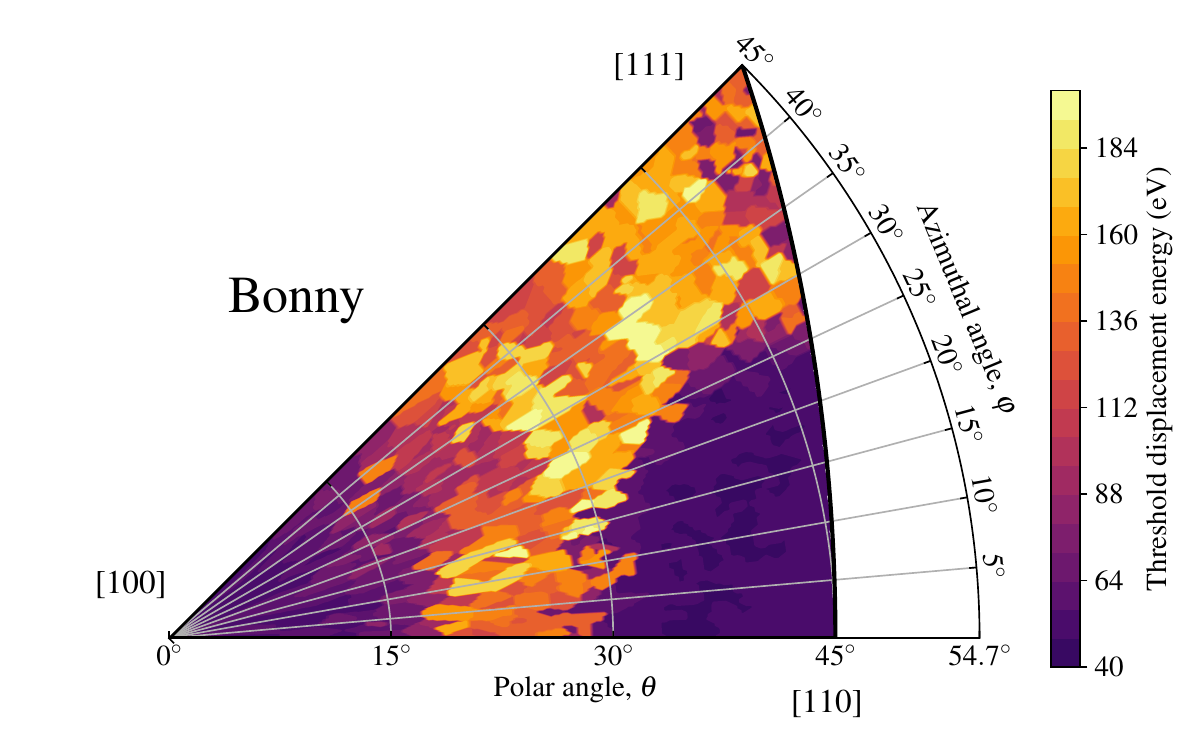}
    \caption{}
\end{subfigure}
    \caption{Threshold displacement energy as a function of PKA direction in Ni. The color range is 10--160 eV for all cases except Bonny which has 10--200 eV. This is due to the significantly higher TDEs in the Bonny potential.}
    \label{fig:tde:Ni}
\end{figure*}

\begin{table*}
 \caption{Threshold displacement energies in Ni given for both high-symmetry directions and the total average TDE. Energies are given in eV and inside the parentheses is the average TDE of all data points that are within a five degree cone from the given specific direction.}
 \label{tab:TDE_Ni_all}
 \begin{threeparttable}
  \begin{tabular*}{\linewidth}{@{\extracolsep{\fill}} llrrrrr }
   \toprule
   &Expt. & Ni tabGAP &  \texttt{Ni\_pv} tabGAP  & Stoller\tnote{a}  & Bonny 2013\tnote{b}  & AIMD\tnote{c,d}\\
   \bottomrule
    \hkl<100> &38\tnote{e}& 21 (23.3 $\pm$ 0.5) & 23 (24.7 $\pm$ 0.4) & 27 (27.9 $\pm$ 0.3) & 49 (52.7 $\pm$ 0.7) & 27\tnote{c}, 24\tnote{d} \\
    \hkl<110> &21\tnote{e}& 19 (19.5 $\pm$ 0.2) & 25 (26.8 $\pm$ 0.2) & 23 (25.2 $\pm$ 0.2) & 49 (51.7 $\pm$ 0.3) & 30\tnote{c}, 17\tnote{d}  \\  
    \hkl<111> &$>$60\tnote{e}& 29 (58.6 $\pm$ 3.7) & 39 (63.4 $\pm$ 4.6) & 35 (37.8 $\pm$ 0.7) & 69 (121.9 $\pm$ 7.5)& 70\tnote{c}, 54\tnote{d} \\
    avg.&33\tnote{f},69\tnote{g}& 42.9  $\pm$ 0.7 & 46.5 $\pm$ 0.9 & 47.7 $\pm$ 0.8 & 99.7 $\pm$ 1.7 & \\
   \bottomrule
  \end{tabular*}
  \begin{tablenotes}
  \item[a] \cite{stoller2016impact}
  \item[b] \cite{bonny2013interatomic}
  \item[c] \cite{yang2021full} 
  \item[d] \cite{liu2015ab} 
  \item[e] \cite{vajda1977anisotropy}
  \item[f] \cite{Luc75}
  \item[g] \cite{jung1981average} 
  \item 
  \end{tablenotes}
 \end{threeparttable}
\end{table*}

We will now turn to actual radiation damage simulations and properties, starting with the threshold displacement energies (TDEs). The threshold displacement energy defines the minimum energy required to displace an atom from its lattice position, creating a stable defect. The TDE is often used as an input parameter for various models, such as when estimating doses with the common NRT equation~\cite{nordlund_primary_2018}. It is important for models that are used in radiation damage simulations to properly describe TDEs, as they have a direct correlation with the amounts of defects produced from cascade events~\cite{nordlund2018improving, NORGETT197550}. 

Fig.~\ref{fig:tde:Ni} shows the TDEs calculated as a function of lattice direction for the different interatomic potentials. All potentials show a darker region of low TDEs close to the \hkl<100> and \hkl<110> directions. Furthermore, the two tabGAPs and Stoller \etal potentials seem reasonably similar, with the Ni\texttt{\_pv} tabGAP and Stoller \etal being most alike. However, the Bonny potential is again a clear outlier, for which we even use a different color scale due to the significantly larger TDE values. Comparing the two tabGAP potentials reveals that the lower TDEs are quite similar, while the higher TDEs are somewhat higher in the Ni\texttt{\_pv} tabGAP. Table~\ref{tab:TDE_Ni_all} shows the TDEs for some high-symmetry directions of the lattice as well as the average TDE over all directions, compared with both experiment and \textit{ab initio} MD calculations. Comparing the two tabGAP potentials the Ni\texttt{\_pv} potential has slightly higher TDEs. However, this difference is surprisingly small. Between the MD potentials there is disagreement between what is the lowest energy direction. However, the difference between \hkl<100> and \hkl<110> in both tabGAP potentials is very small and the same order as the 2 eV energy increment used in the simulations. Both tabGAP potentials and the Stoller potentials all place the average TDEs somewhere between 40 and 50 eV, with Stoller and Ni\texttt{\_pv} tabGAP being remarkably close to each other. The Bonny potential predicts significantly higher TDEs on the other hand.

The ASTM recommends~\cite{ASTM} 40 eV as the effective TDE for Ni. On the other hand, Yang \etal proposes 70 eV as a more reasonable estimate of the true average~\cite{yang2021full}. However, the results in this work suggest that the average is likely closer to the ASTM average than 70 eV. Now the question arises, are the results presented here reasonable? At least methodologically, the approaches should be quite similar as the average TDE we report for the Bonny potential is precisely the same as Yang \etal report for the same potential. Unfortunately, there is a large spread in the available experimental estimates of the average TDEs. Currently, there exist some discrepancies in the reported literature both from a wide range of classical potentials and between MLIPs and AIMD, that new experimental investigations into the threshold displacement energies in Ni should be carried out. Rather surprisingly, the two different ML potentials (with different pseudopotentials), report very similar TDEs in Ni. However, this does not mean that this should be the case for other materials. Based on Fig.~\ref{fig:trajectory_Ni} the are significant differences between the pseudopotentials that could play a role in other materials.    

\subsection{Primary radiation damage}

\begin{figure}
    \centering
    \includegraphics[width=0.99\linewidth]{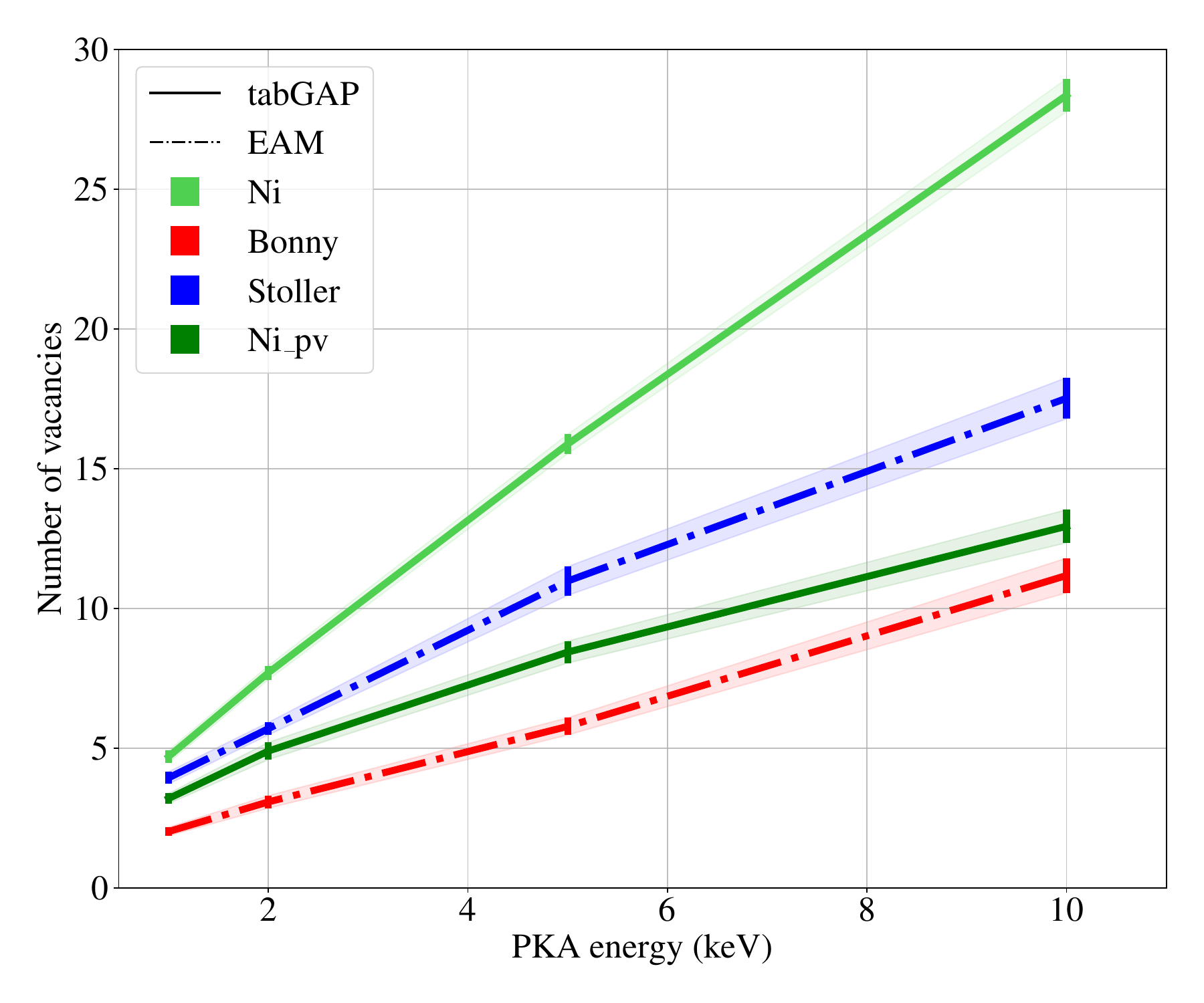}
    \caption{The average number of produced vacancies (from WS analysis) from single cascades with different PKA energy with different Ni potentials. Shaded regions indicate standard deviation of 50 individual cascades.}
    \label{fig:PKA-def}
\end{figure}

Primary radiation damage in Ni has been studied extensively throughout the years using MD \cite{nordlund_defect_1998,Nor98,granberg_mechanism_2016,yao2007study,nair2025statistical}. Thus, we will not attempt an exhaustive analysis of primary damage here. Like before we will focus on key aspects and highlight the differences between the potentials that can be observed. Fig.~\ref{fig:PKA-def} shows the number of produced vacancies (i.e. Frenkel pairs) as a function of PKA energy. We observe clear differences between the different potentials, with the Ni tabGAP having the highest defect production. The are two aspects of these results that are striking. Firstly, even though the TDEs of the two tabGAP potentials are quite similar the primary damage production is significantly different. This is due to the differences in the repulsive interactions. Secondly, the ordering of the potentials does not follow the order of the average TDEs reported in section~\ref{sec:tde}. The lack of correlation between TDE and displacement damage is a well known phenomenon that has been reported before in the literature~\cite{malerba_molecular_2006}.

\begin{figure}
    \centering
    \includegraphics[width=0.99\linewidth]{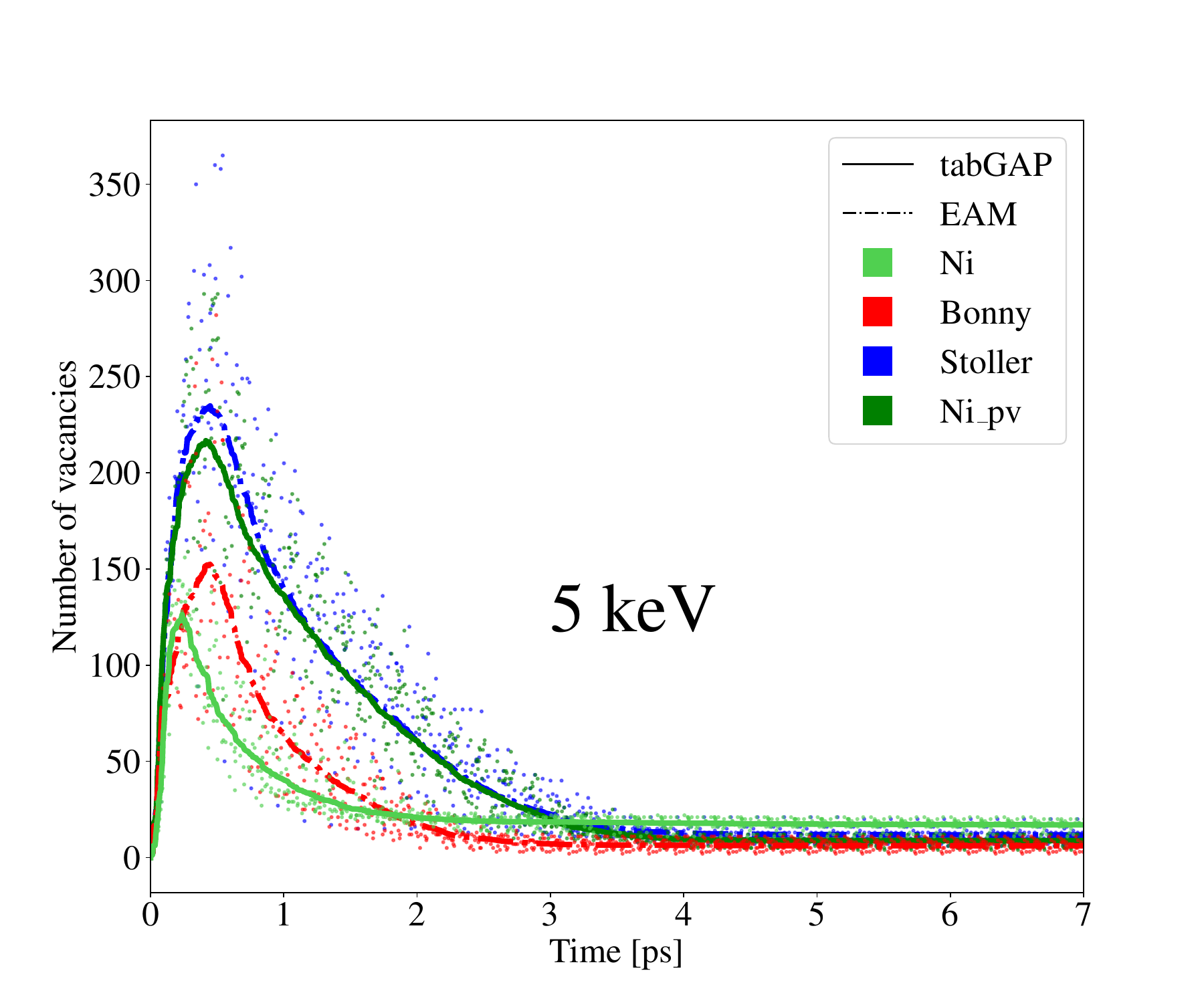}
    \caption{Number of produced vacancies (from WS analysis) during single cascades as a function of time (5 keV PKA energy) in different Ni potentials. The curves shown are rolling averages of all the individual cascades together.}
    \label{fig:typ-casc}
\end{figure}

Fig.~\ref{fig:typ-casc} illustrates typical time-evolutions of 5 keV PKA cascades. Here all the individual cascades have been superimposed and the curves indicate rolling averages of all the individual cascades together. Here, we get a completely different point of view, where the Ni tabGAP's peak damage is clearly the lowest, Ni\texttt{\_pv} and Stoller are high and quite similar, and Bonny is somewhere in the middle. Additionally, we notice that the peak damage does not determine the ordering at the end. Looking at the individual cascades we find that the Ni\texttt{\_pv} tabGAP typically produces denser cascades than the Ni tabGAP potential. This is in line with what has been reported for tungsten~\cite{sand_non-equilibrium_2016,de2021modelling}, where it was noted that softer potential produces weaker heat-spikes and larger (and less spherical) cascades which produces more defects. 

\subsection{Overlapping cascades}

\begin{figure}
    \centering
    \includegraphics[width=0.99\linewidth]{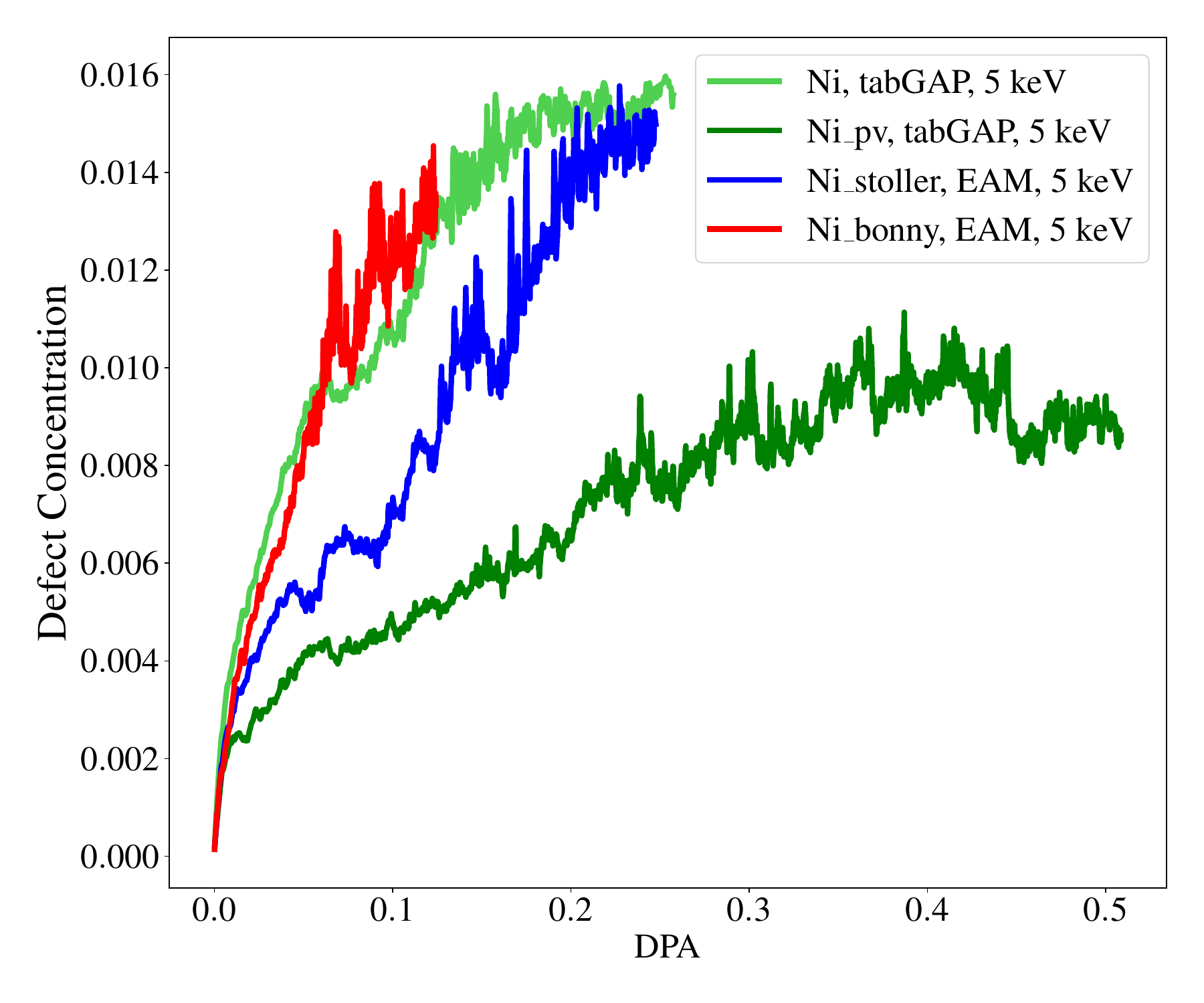}
    \caption{Defect concentration as a function of estimated dose (NRT dpa) for the different Ni potentials. 2000 cascades for Ni tabGAP, Stoller and Bonny potentials. 4000 cascades for Ni\texttt{\_pv} tabGAP potential.}
    \label{fig:defects-dpa}
\end{figure}

\begin{figure*}
    \centering
\begin{subfigure}[t]{0.49\textwidth}
\label{fig:ff:nitabgap}
    \includegraphics[width=\textwidth]{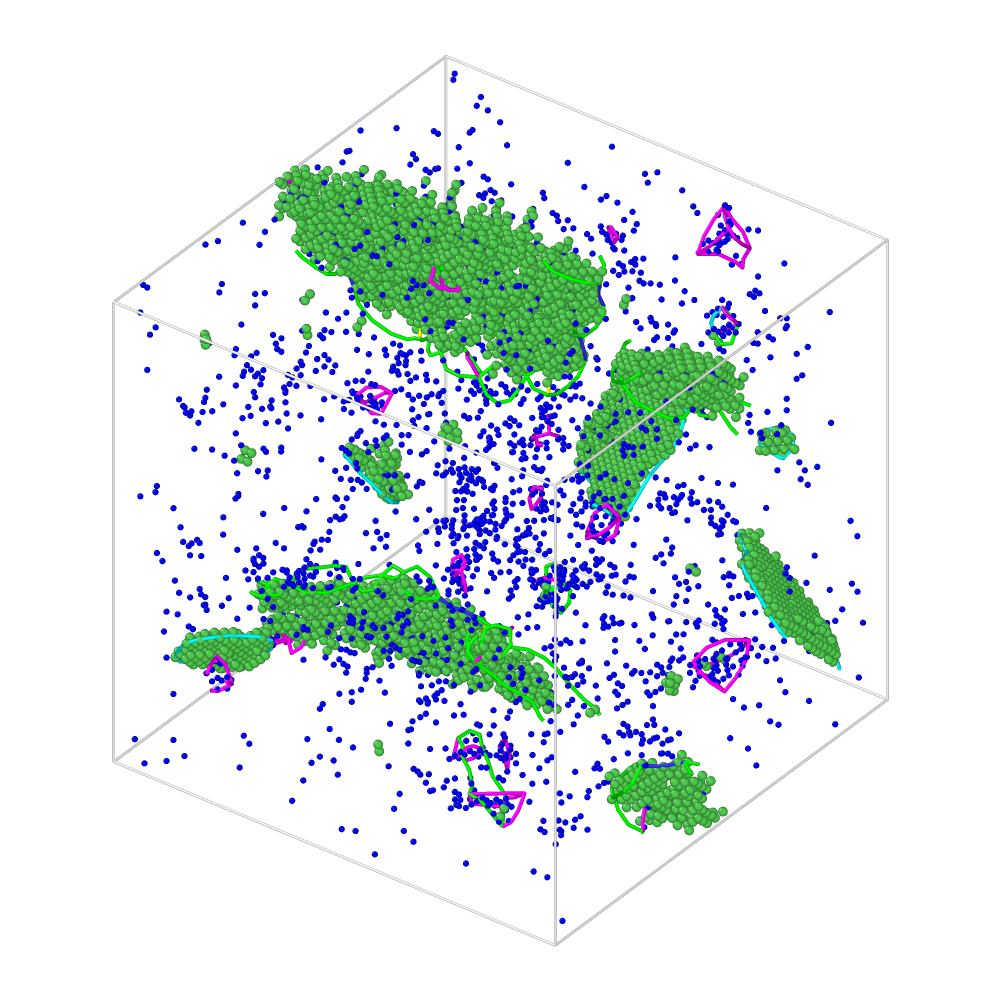}
      \caption{Ni tabGAP}
\end{subfigure}
\begin{subfigure}[t]{0.49\textwidth}
\label{fig:fg:nipvtabgap}
    \includegraphics[width=\textwidth]{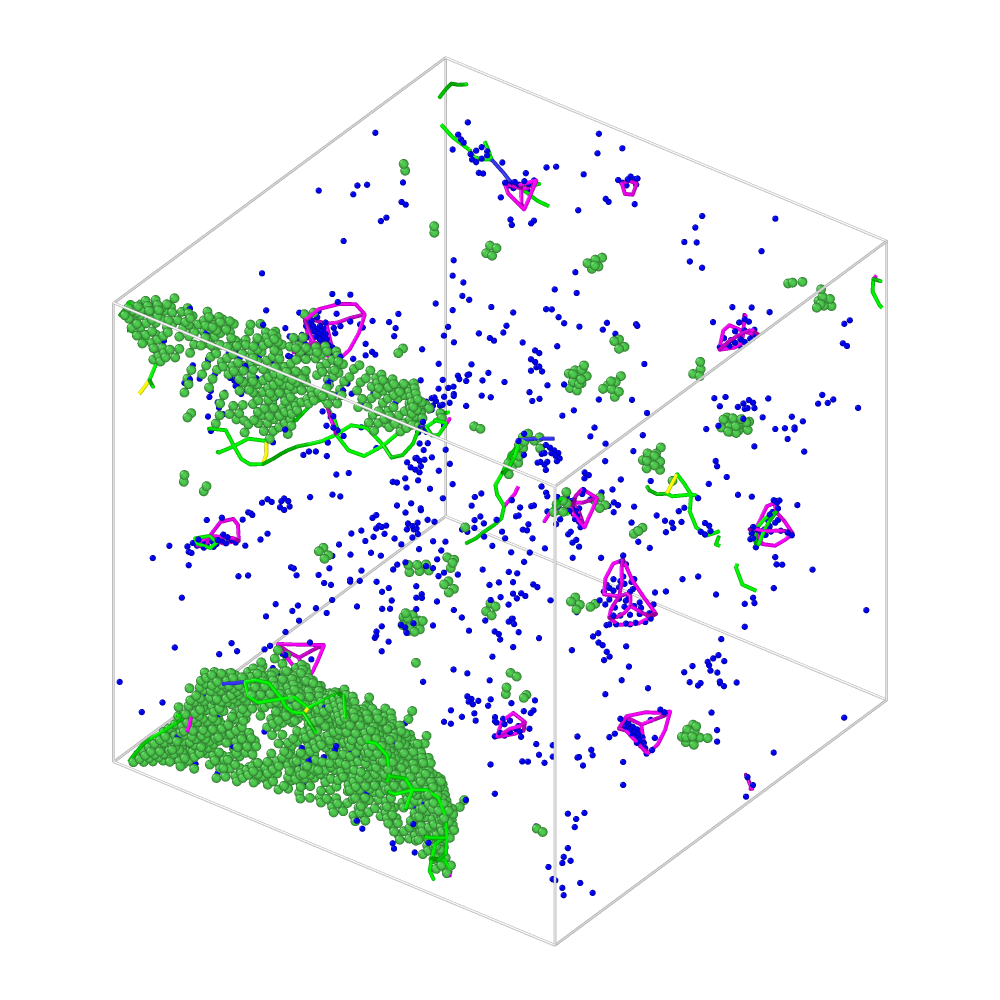}
    \caption{Ni\texttt{\_pv} tabGAP}
\end{subfigure}
\begin{subfigure}[t]{0.49\textwidth}
\label{fig:ff:stoller}
    \includegraphics[width=\textwidth]{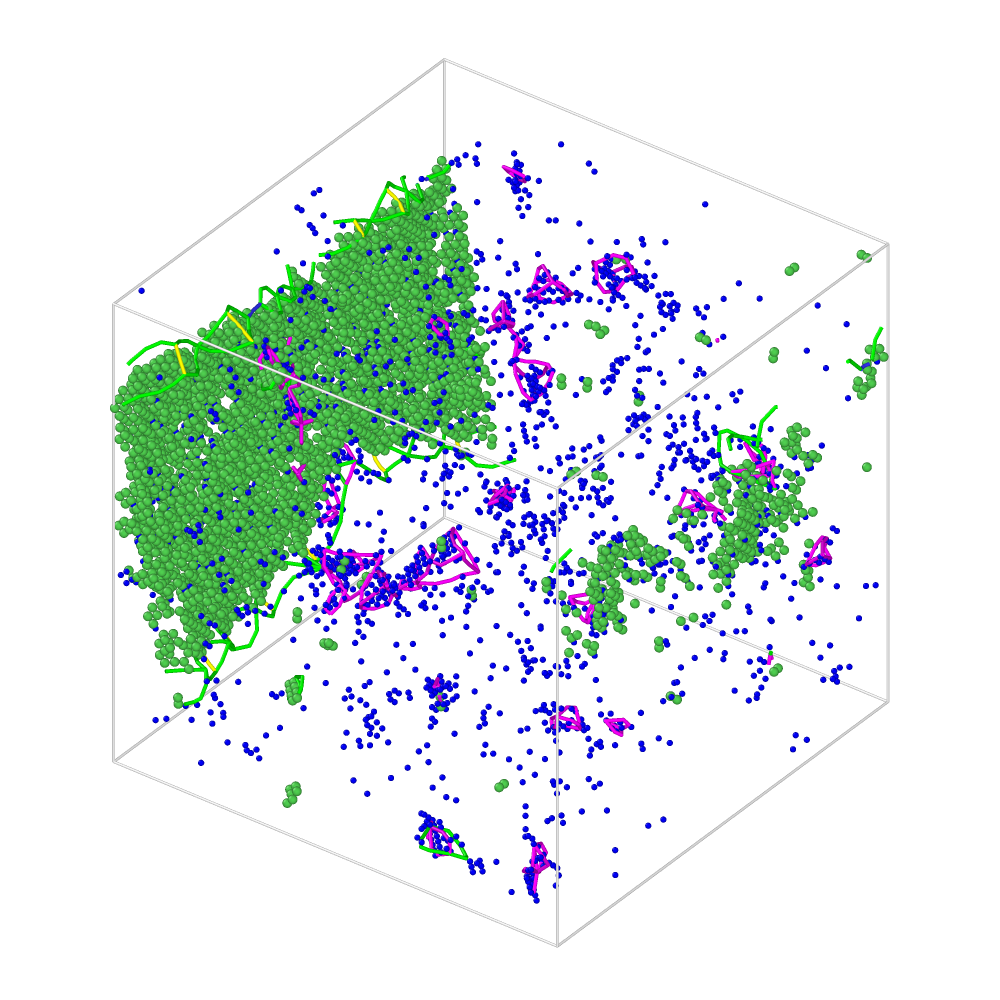}
    \caption{Stoller EAM}
\end{subfigure}
\begin{subfigure}[t]{0.49\textwidth}
\label{fig:ff:bonny}
    \includegraphics[width=\textwidth]{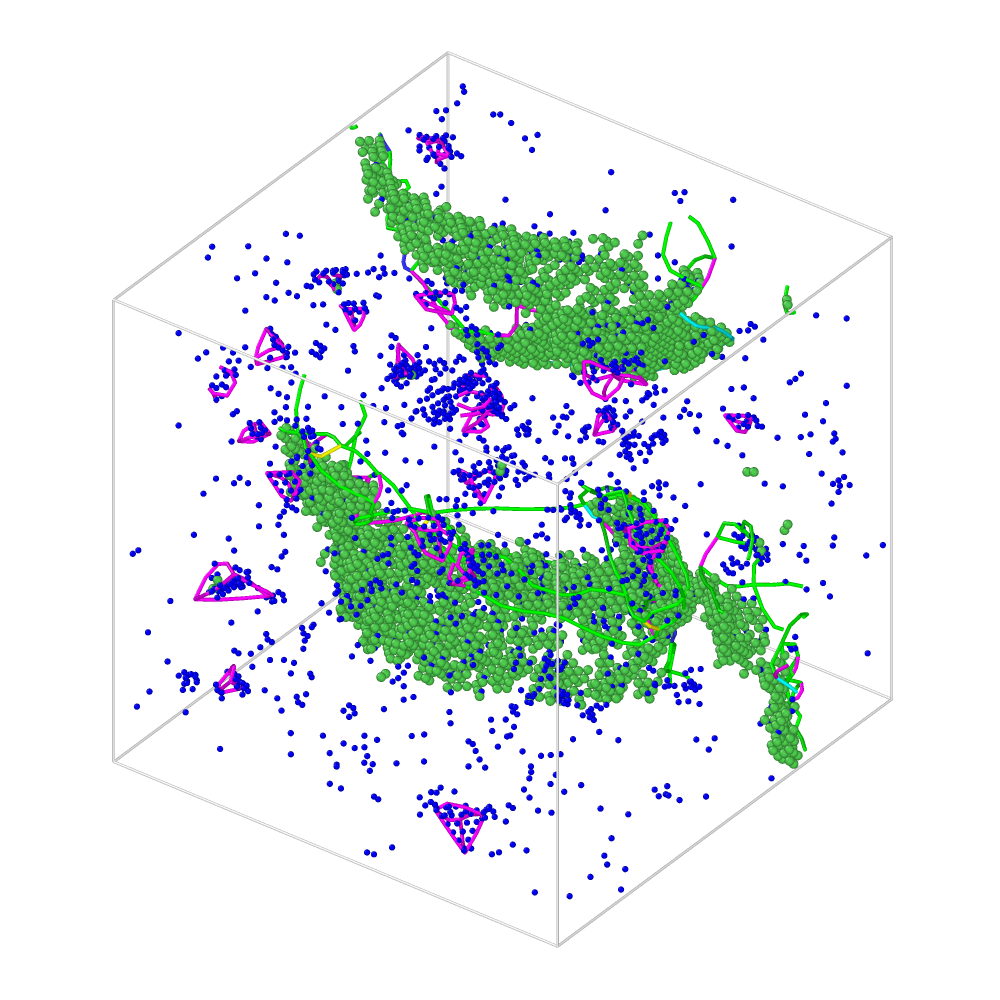}
    \caption{Bonny EAM}
\end{subfigure}
    \caption{Final defect configurations after 2000 overlapping collision cascades. Blue spheres represent vacancies, Green spheres represent interstitial atoms, green lines represent Shockley dislocations, and cyan lines represent Frank dislocations and pink lines represent stair-rod dislocations.}
    \label{fig:ff:Ni}
\end{figure*}

This far, we have only looked at individual displacement cascade events. However, in reality during continued irradiation it becomes increasingly likely that collision cascades are produced in the vicinity of pre-existing damage. There exist a number of MD studies on the effects of cascade overlap both in the case of well defined defects and on displacement damage~\cite{fellman_radiation_2019,fellman2022recoil,wang_molecular_2019,sand_defect_2018,granberg_cascade_2017,byggmastar_collision_2019}. Furthermore, massively overlapping cascades have also been studied in Ni, where it was noticed that there is significant differences between different potentials~\cite{levo_radiation_2017,levo2021temperature}. Fig.~\ref{fig:defects-dpa} shows the defect concentration as a function of the number of cascades. Here we see something striking: all the potentials except the Ni\texttt{\_pv} tabGAP seem to converge to the same defect concentration and the difference to Ni\texttt{\_pv} tabGAP is significant. This was indeed so surprising that we continued the simulations in the Ni\texttt{\_pv} potential for an additional 2000 cascades, with the assumption that it would at some point converge with the rest. However, this is not the case. Fig.~\ref{fig:ff:Ni} shows the defect structures of the different potentials after 2000 cascades. At first glance, all the defect structures look qualitatively similar. In all potentials the simulation cell is dominated by a large interstitial-type dislocation formed from Shockley partial chains. Furthermore, we see in all potentials stacking fault tetrahedra (SFTs) made up of vacancies, but the amount of these differs between the potentials. We also clearly see a lower defect concentration in the Ni\texttt{\_pv} potential compared to the rest. The final defect structure of the Ni\texttt{\_pv} potential after 4000 cascades can be found in the supplementary materials. However, it is qualitatively similar to the one after 2000 cascades. 

\begin{figure}
    \centering
    \includegraphics[width=0.99\linewidth]{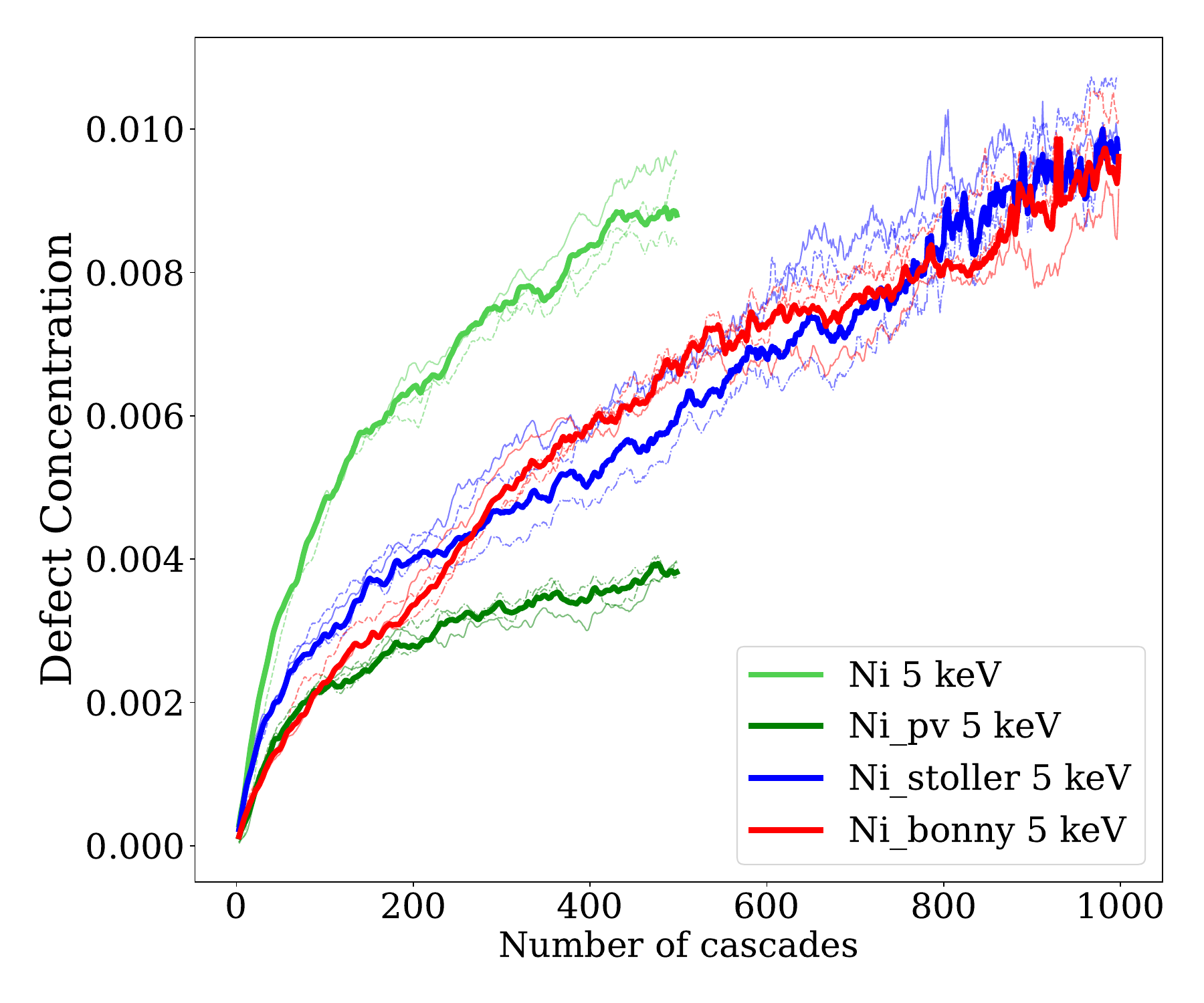}
    \caption{Defect concentration as a function of number of collision cascades in the different Ni potentials. Faint curves are the individual runs and the bold curve is the average.}
    \label{fig:short-runs}
\end{figure}

Now we can ask whether the defect structure we get at the end is completely random, meaning that with different initial values we could get completely different end results. Looking at results from literature, this does not seem to be the case~\cite{GRANBERG2017114}. Nevertheless, in order to verify this we re-ran the collision cascades with 3 random initializations for several hundred cascades (500 tabGAP and 1000 EAM) to see if there are massive deviations between the different runs. The results of this can be seen in Fig.~\ref{fig:short-runs} and we can conclude that while there is some random variation, it is not significant enough to change the conclusions.

The difference between the two tabGAPs is somewhat expected, as there are significant differences already in the single cascades. Additionally, the fact that different EAM potentials give different results is also something that has been reported previously, and to be expected. However, what is surprising is the fact that the Stoller and  Ni\texttt{\_pv} potentials give so different results as up until this point they seemed to behave most alike in the repulsive interactions, referring now to the dimer curves, QSD, TDEs and single cascades presented earlier. There are numerous implications of this. First, this implies that just because the machine-learning potential reproduces equilibrium properties well and has a ZBL (or some other) repulsive potential, it does not guarantee similar results if the transition to the repulsive part is not properly defined. This is rather obvious but deserves re-stating. Second, just because the EAM potential behaves very similarly in the case of repulsive interactions to DFT or MLIPs, like Stoller and Ni\texttt{\_pv} in our case, does not guarantee similar results for cumulative cascade damage either. Now the unfortunate (and obvious) conclusion becomes that we need both the repulsive and equilibrium interactions to be accurately represented in the interatomic potential. Another, equally unfortunate conclusion is that validating an interatomic potential for massively overlapping cascade simulations remains challenging, as we cannot just look at repulsive interactions, TDEs and formation energies of defects and expect the same results. 

%
%

\subsection{RBS/c simulations}

\begin{figure}
    \centering
    \includegraphics[width=0.99\linewidth]{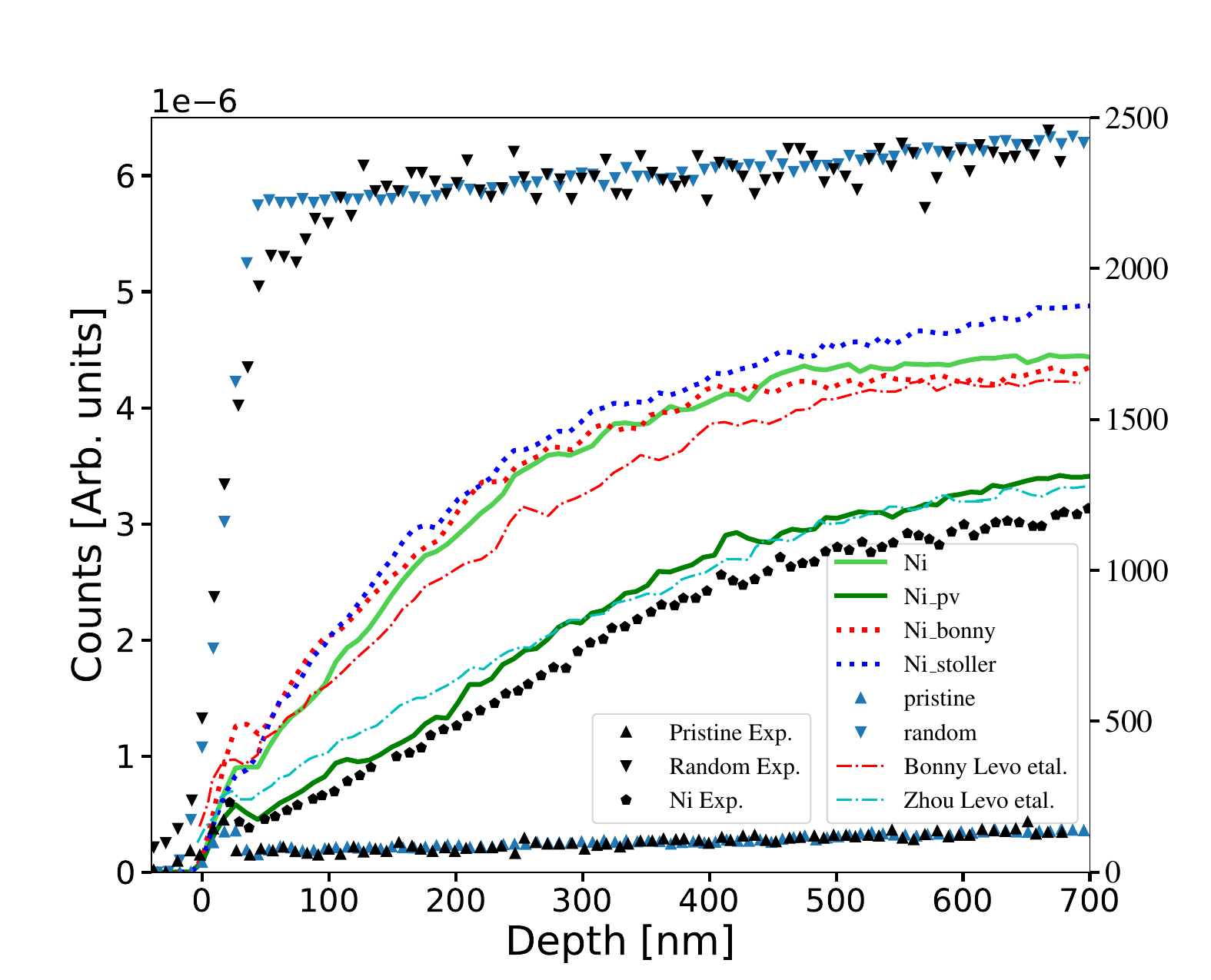}
    \caption{Comparison between simulated and experimental RBS/c spectra at 300K. Experimental results from Ref.~\cite{Zha15b} and simulated spectra for Bonny and Zhou potentials from Levo \etal~\cite{levo2021temperature}. Simulated structures sampled with assumption of TDE of 40 eV and no electronic loss.}
    \label{fig:rbs-tde-40}
\end{figure}

The experimental validation of the results of the overlapping cascade simulation remains challenging. This is due to the inherent challenges in making direct comparisons with experiments and the lack of long-term defect evolution in MD. One approach that can give direct comparison with experiments is the simulation of RBS/c spectra \cite{zhang2016simulation,jin2020new}. Previously, simulated RBS/c spectra has been used for both Ni and Ni-based alloys where valuable comparisons have been achieved \cite{zhang2017radiation,levo2021temperature}. In Fig.~\ref{fig:rbs-tde-40} we show the simulated RBS/c spectra of the different potentials compared with experimental results. The spectra in Fig.~\ref{fig:rbs-tde-40} follow closely the procedure detailed in Ref.~\cite{levo2021temperature}. In the figure, we have scaled the y-axis of the simulated spectra so that the simulated pristine and random samples are of similar magnitude. Comparing the results for the Bonny potential from this work and that reported in Ref.~\cite{levo2021temperature}, we get very similar but not identical results. Slight differences are to be expected as the source structures are not the same, however this demonstrates that the methodology has been faithfully reproduced. If we now look at the different potentials we see a few interesting findings. First, the Ni\texttt{\_pv} potential is clearly different from the rest, closest to the experimental results and quite similar to the spectra that Levo \etal report for the Zhou \etal potential~\cite{zhou2004misfit}. Second, rather surprising is the fact that the Stoller \etal potential show the highest yield out of all the simulated spectra. Looking at Fig.~\ref{fig:defects-dpa} from which the RBS/c structures have been sampled from, the Stoller  potential had defect concentrations in between the other potentials.

\begin{figure}
    \centering
    \includegraphics[width=0.99\linewidth]{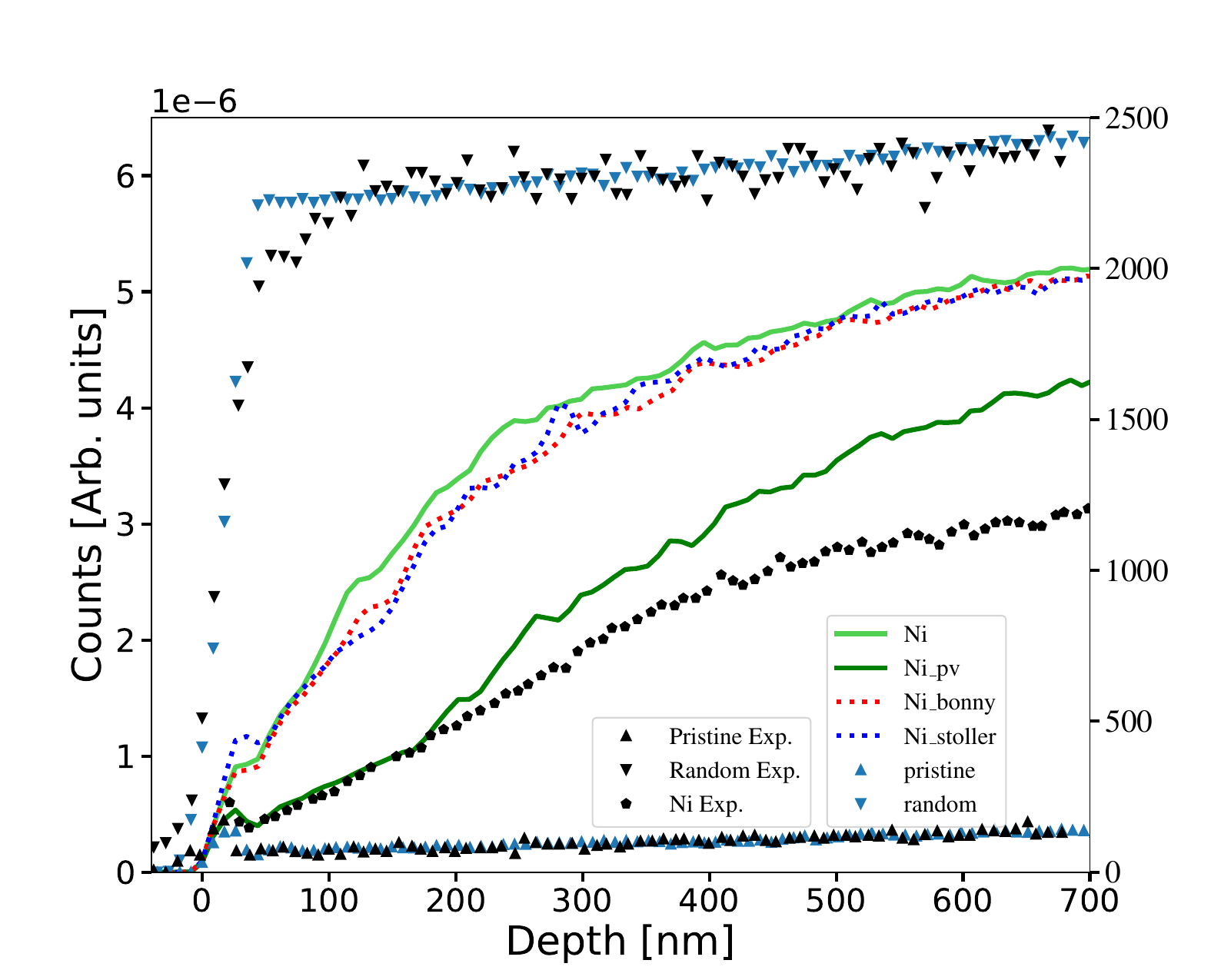}
    \caption{Comparison between simulated and experimental RBS/c spectra at 300K. Experimental results from Ref.~\cite{Zha15b}. Simulated structures sampled using nuclear deposited energy profile show in Fig~\ref{fig:depens}.}
    \label{fig:rbs-depen}
\end{figure}

In the simulated spectra shown in Fig.~\ref{fig:rbs-tde-40} and in the work by Levo \etal, there are a few assumptions and approximations underlying them. In this case the structures are sampled using a nuclear energy deposition profile using a fluency half of the experimental one. Furthermore, the sampling uses as estimation of dose based on the NRT dpa equation with the assumption of 40 eV TDE and presumably no electronic loss. As we have seen in section~\ref{sec:tde}, there are meaningful differences in the predicted TDEs of the potentials (especially between Bonny and the rest). Hence we would like to avoid making such assumptions of the TDE. To address this, we redid the calculations of the nuclear energy deposition profile in order to more closely match the experimental conditions. Additionally, by sampling the structures directly from the nuclear deposited energy, we avoid making any assumption of the TDE. Fig~\ref{fig:rbs-depen} shows the simulated spectra with the new sampling methodology (i.e. from Fig.~\ref{fig:depens}). Unsurprisingly, due to the increased nuclear deposited energy (thus dose) the simulated yields are somewhat higher. Compared to the spectra in Fig.~\ref{fig:rbs-tde-40} the comparison with experimental results is somewhat worse, but the overall conclusion does not change, which is that out of the different models the Ni\texttt{\_pv} potential predicts defects structures closest to the experimental results.

\subsection{Modifications to the repulsive pair interactions}\label{sec:res:mod}

\begin{figure}
    \centering
    \includegraphics[width=0.99\linewidth]{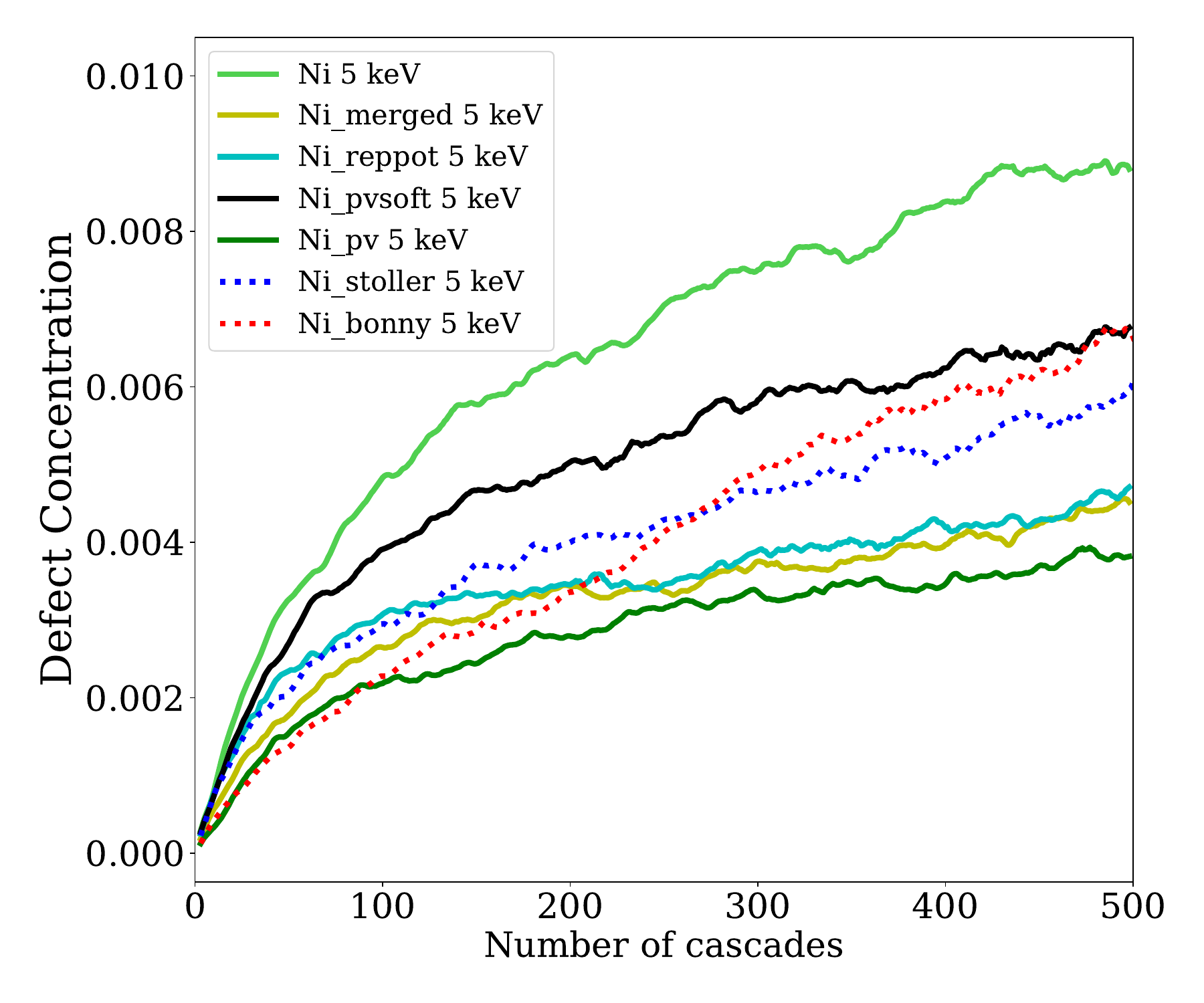}
    \caption{Defect concentration as a function of number of collision cascades in the different Ni potentials, including potentials with modified two-body terms.  The curve are averages of three runs. Ni$_{\mathrm{merged}}$ refers to the ''soft'' potential where the two-body term is smoothly transitioned to the Ni\texttt{\_pv} potential at short distances. Ni\texttt{\_pvsoft} is the opposite where the Ni\texttt{\_pv} potentials two-body term is transitioned to the ''soft'' potential at short-distances. Ni$_{\mathrm{reppot}}$ refers to ''soft'' potential where the two-body term has been refitted to the original external repulsive potential. }
    \label{fig:modified-2b}
\end{figure}

In the previous sections, we have shown the differences between the MLIPs in regards to radiation damage simulations. We have demonstrated the need for a proper description of both short-range and equilibrium interactions. In this section we look at the possible approaches to modifying the of the MLIPs to improve the repulsive interactions. We are fortunate that the tabGAP formalism gives us direct access to the pair interactions, making modifications easy to implement. We do this due to two reasons; First, to get a sense of the validity of such approaches and secondly to investigate what are the implications of modifying just the repulsive pair interactions of the potential. The different modifications have been detailed in section~\ref{sec:met:mod}, but as a reminder, the Ni$_{\mathrm{reppot}}$ is the potential where the ''soft'' Ni tabGAP potential is refitted with its own external repulsive potential, Ni$_{\mathrm{merged}}$ is the potential for which the two-body terms have been merged in a way that the Ni\texttt{\_pv} constitute the short-range pair-interactions and the Ni\texttt{\_pvsoft} is the merged so as to ''soften'' the potential by merging it to the ''soft'' Ni tabGAP at short distances. The effects of this merging can be clearly seen in the additional dimer curves provided in the supplementary materials. Using these new potentials, we simulated the overlapping cascades again. Fig.~\ref{fig:modified-2b} shows the results of these calculations comparing against the previous results. We can immediately see that the modifications which work to ''harden'' the Ni tabGAP potential have significantly reduced the defect concentrations, bringing these results surprisingly close to the unmodified Ni\texttt{\_pv} tabGAP potential. Conversely, ''softening'' the  Ni\texttt{\_pv} potential significantly increases its predicted defect concentration. However, in this case it is still significantly lower than the unmodified Ni tabGAP potential. 

These result imply that indeed these methods could be used to improve the potential after its initial training either by refitting the repulsion (''hardening'') or by swapping between two different models. However, the modified potentials in this work should not be used in production runs as they have not been validated in any other respects. Furthermore, the joining procedure was done in a naive way by just looking at the dimer curves and quasi-static drag by eye and thus a more sophisticated optimization procedure of the joining is advisable. Nonetheless, these results both demonstrate that it is feasible to make such modifications. However, it goes without saying that such modifications are somewhat artificial and as such should be carefully validated in each case. Furthermore, these results confirm the assumption that the differences between the two tabGAP potentials is mainly due to the differences in the repulsive pair interactions.  In a study on iron by Byggmästar \etal, it was concluded that modifications to the repulsive interactions did not have a large impact on the final accumulated defect concentration~\cite{byggmastar_effects_2018}. It was further noted that final produced defects in overlapping cascade simulations is mainly controlled by the stability and mobility of defect clusters defined by the near-equilibrium part of the potential. However, the results from this work suggest that this might not be true in general, and is most likely dependent on both the material and the energy range in which the repulsive interactions differ.

\section{Conclusions}

In this work, we have looked in detail at the differences between machine-learned interatomic potentials trained to DFT data computed using two different pseudopotentials (Ni and semicore Ni\texttt{\_pv}). Both potentials were tabulated Gaussian approximation potentials. We have focused on the aspects related to simulation of radiation damage in particular. 

We found that dimer curves and quasi-static drag (QSD) curves remains a reasonable check for the potentials ability to describe short-range interactions. However, significant differences can be found in the results despite similar dimer curves and QSD. The short-range interactions can be tricky to get right during the training of the potential and significant differences can be found between the MLIPs.  There were minimal differences between the MLIPs with regards to equilibrium properties. Even though there were significant differences in the dimer curves and QSD, the average threshold displacement energies were quite similar between the MLIPs (Ni tabGAP 43 eV and Ni\texttt{\_pv} tabGAP 47 eV).  Overall the results give support to an average TDE between 40-50 eV for nickel. 
 
 Additionally, there are significant differences with regards to primary damage and typical cascade characteristics between the potentials.  In massively overlapping cascade simulations, the Ni\texttt{\_pv} tabGAP potential showed clearly lower defect concentrations to the rest of the potentials. Interestingly, the Stoller \etal EAM potential which in most other respects showed very similar behavior to the Ni\texttt{\_pv} tabGAP potential, showed drastically different (higher) defect concentration compared to the Ni\texttt{\_pv} potential after continued irradiation.
    
We also made direct comparisons to experiments of the overlapping cascades by means of RBS/c simulations. The Ni\texttt{\_pv} tabGAP potential had RBS/c spectra closest to the experimental results, while all other potentials showed significantly higher yields. Furthermore, we discuss the impact of different sampling procedures and the use of different nuclear energy deposition profiles.

Furthermore, we tried out different approaches to modify the repulsive pair interactions of the MLIPs after they have already been trained and found the approaches feasible, while being somewhat ad-hoc. Furthermore, we conclude that  these modifications had significant impacts on the results and could be used to bridge the differences between the MLIPs.

\section*{Acknowledgments}

This work has received funding from the Academy of Finland through the HEADFORE project (grant number 333225).
The authors wish to thank the Finnish Computing Competence Infrastructure (FCCI) and CSC -- IT Center for Science for supporting this project with computational and data storage resources.
This work has been partially carried out within the framework of the EUROfusion Consortium, funded by the European Union via the Euratom Research and Training Programme (Grant Agreement No 101052200 — EUROfusion). Views and opinions expressed are however those of the author(s) only and do not necessarily reflect those of the European Union or the European Commission. Neither the European Union nor the European Commission can be held responsible for them. A. Fellman would like to acknowledge E. Levo for providing the original RBSADEC input files for the RBS/c calculations from the earlier work by Levo \etal.

\bibliography{main}

\end{document}